\documentclass[prx,reprint,amsmath,amssymb,aps,floatfix,dvipsnames,longbibliography]{revtex4-2}
\usepackage[normalem]{ulem}
\usepackage{subfiles}
\usepackage{natbib}
\usepackage{graphicx}
\usepackage{dcolumn}
\usepackage{bm}
\usepackage{hyperref}
\usepackage{xcolor}

\usepackage[normalem]{ulem}

\newcommand{\rjoint}[1]{\textcolor{Blue}{#1}}
\newcommand{\rsplit}[1]{\textcolor{Red}{#1}}

\begin{document}

\preprint{APS/123-QED}

\title{R\'enyi entropy of quantum anharmonic chain at non-zero temperature}

\author{Miha Srdinšek\textsuperscript{1,2,3}}
\email{miha.srdinsek@upmc.fr}
\author{Michele Casula\textsuperscript{2}}
\author{Rodolphe Vuilleumier\textsuperscript{3}}
\affiliation{
 \textsuperscript{1}Institut des sciences du calcul et des donn\'ees (ISCD), Sorbonne Universit\'e, 4 Place Jussieu, 75005 Paris.\\
 \textsuperscript{2}Sorbonne Universit\'e, Institut de min\'eralogie, de physique des mat\'eriaux et de cosmochimie (IMPMC),  CNRS UMR 7590, MNHM, 4 Place Jussieu, 75005 Paris.\\
 \textsuperscript{3}Processus d’Activation S\'electif par Transfert d’Energie Uni-\'electronique ou Radiative (PASTEUR), CNRS UMR 8640, D\'epartement de Chimie, \'Ecole Normale Superieure, 24 rue Lhomond, 75005 Paris.
 }

\date{\today}

\begin{abstract}
  The interplay of quantum and classical fluctuations in the vicinity of a quantum critical point (QCP) gives rise to various regimes or phases with distinct quantum character. In this work, we show that the R\'enyi entropy is a precious tool to characterize the phase diagram of critical systems not only around the QCP but also away from it, thanks to its capability to detect the emergence of local moments at finite temperature. For an efficient evaluation of the R\'enyi entropy, we introduce a new algorithm based on a path integral Langevin dynamics combined with a previously proposed thermodynamic integration method built on regularized paths. We apply this framework to study the critical behavior of a linear chain of anharmonic oscillators, a particular realization of the $\phi^4$ model. We fully resolved its phase diagram, as a function of both temperature and interaction strength. At finite temperature, we find a sequence of three regimes - para, disordered and antiferro-, met as the interaction is increased. The R\'enyi entropy divergence coincides with the crossover between the para and disordered regime, which shows no temperature dependence. The occurrence of the antiferro regime, on the other hand, is temperature dependent. The two crossover lines merge in proximity of the QCP, at zero temperature, where the R\'enyi entropy is sharply peaked. Via its subsystem-size scaling, we confirm that the transition belongs to the two-dimensional Ising universality class. This phenomenology is expected to happen in all $\phi^4$-like systems, as well as in the elusive water ice transition across phases VII, VIII and X.
\end{abstract}

\maketitle

\section{Introduction}
\label{sec:introduction}

Classical phase transitions are driven by thermal fluctuations, which affect the volume of explored phase space. Due to the wave nature of quantum mechanics, fluctuations appear even at zero temperature, this time stemming from the uncertainty principle of wave mechanics. At sufficiently low temperature, the interplay of both effects results in a rich phase diagram\cite{Vojta2003,Sachdev2008,Sachdev2011}. Quantum fluctuations typically suppress phase transitions, and direct the classical critical region towards the quantum critical point (QCP) at lower temperatures. This effect has been experimentally observed, for instance, in  the Ising-like transition in LiHoF$_4$\cite{Bitko1996}, in the so-called quantum para-electric materials (SrTiO$_3$, BaTiO$_3$, KTaO$_3$), in quantum dielectric compounds, such as $\kappa$-ET$_2$Cu$_2$(CN)$_3$\cite{Abdel-Jawad2012} and $\kappa$-ET$_2$Cu[N(CN)$_2$]Cl\cite{Lunkenheimer2012}, and in structural phase transitions of hydrogen-bonded materials, such as water ice VII-VIII-X\cite{Pruzan2003}, superconducting hydrides LaH$_{10}$\cite{Drozdov2019}, YH$_n$\cite{Kong2021} and H$_3$S\cite{Drozdov2015}, and hydrogen halides, like HF and HBr\cite{Jansen1987,Springborg1988,Wang1994}. Throughout their phase diagram, the interplay of quantum and thermal fluctuations introduces typical regimes with distinct scaling laws and entanglement properties\cite{Gabbrielli2018,Lu2020,Frerot2019}. A prominent example of such regimes is the quantum critical behavior above the QCP, where long-range order is destroyed while local moments, like spin or local polarisation can be preserved, locally breaking a symmetry. A prototypical example of this situation is  found in the one-dimensional (1D) transverse Ising model, where three distinct regimes are well understood to be a consequence of different types of symmetry-breaking mechanisms\cite{Sachdev1997, Wu2020}.

All previously mentioned experimental systems can be approximated by a generalisation of the Ising model - the discretised $\phi^4$ model -, where particles live in a double well external potential, interacting through a quadratic term. By increasing the interaction, they freeze in a long-range ordered configuration through a phase transition belonging to the Ising universality class\cite{Toral1990,Rubtsov2001,Kim2007}. The transition can be of displacive (soft mode) or order-disorder type, depending on the height of the double-well barrier. When the barrier is low, particles fluctuate around the origin until the interaction displaces them, by softening their shuttling vibrational mode. Conversely, in the large barrier regime, particles tunnel or switch between off-centered positions and, eventually, their interaction can force them to occupy the same double-well minimum if the coupling is ferro, or the opposite one in an antiferro model. 

The finite-temperature phase diagram of the $\phi^4$ model continues to be a field of active research, as it can change considerably from system to system\cite{Hotta2022}. In higher dimensions, the phase diagram shows the occurrence of classical phase transition, induced by thermal fluctuations, emerging from quantum criticality\cite{Savkin2002}. However, different regimes surrounding the QCP are not well understood, with the exception of the infinite-barrier limit, which is analogous to the Ising model. This owes mostly to the fact that not all the physical mechanisms associated with quantum criticality are known and, hence, reliable physical and experimental probes are hard to devise. As a result, phase diagrams of many critical systems, such as certain super-hydrides\cite{Benoit1998,Bronstein2014,Cherubini2021,Pruzan2003,Reinhardt2022} or the superconducting cuprates\cite{Orenstein2000,simon2002detection,varma2006theory,Lee2008, kim2008theory, Sachdev2010,Kowalski2021} as notable examples, are not fully explained yet.

On the other hand, entanglement and entropic properties have proven to be valuable tools for the analysis and classification of quantum phase transitions in many-body systems\cite{Vidal2003,Cirac2008}. One of the most commonly used measures is the R\'enyi entropy, as it directly quantifies quantum and classical fluctuations\cite{Calabrese2004,Hastings2010,Humeniuk2012,Alba_2013,Alba2017,Demidio2020,DEmidio2022,Tubman2013,Luitz2014,Buividovich2008,Zhao2022new,Zhao2022,Yingfei2017,Bertini2022,Calabrese2020}. Its scaling with the subsystem size is markedly different at the QCP\cite{Eisert2010}. Particularly in 1D systems, the entanglement of a pure state generally saturates at some finite system size, while at the QCP it diverges as a logarithm. The prefactor in front of the logarithm can be analytically shown to be equal to the central charge of the conformal field theory describing the QCP, one of the most striking features of 1D systems\cite{Vidal2003,Calabrese2004}. 

R\'enyi entropy allows to detect the emergence of a local symmetry breaking and the corresponding local moment formation at finite temperature. In this work, we consider the R\'enyi entropy to characterize the phase diagram not only around the QCP but also away from it. With the use of the replica trick ideas in path integral simulations\cite{Hastings2010,Humeniuk2012,Alba_2013,Alba2017,Demidio2020,DEmidio2022,Tubman2013,Luitz2014,Buividovich2008,Zhao2022new,Zhao2022}, the R\'enyi entropy can be evaluated at finite temperature for a targeted subsystem. However, the R\'enyi entropy evaluation is very computationally demanding, with errors increasing with system size, level of entanglement and area of subsystem boundaries. With the aim at making the R\'enyi entropy evaluation much more easily accessible, we introduce a new algorithm based on path integral equations, sampled by a Langevin dynamics, that exploits a recently developed thermodynamic integration method\cite{Srdinsek2021}. 

We then present a full study of the low-temperature phase diagram of the 1D discrete $\phi^4$ model in the regime of vanishing double-well barrier, where it represents a chain of antiferromagnetically coupled an-harmonic oscillators. By computing the R\'enyi entropy, we accurately predict the QCP location. We perform the subsystem-size scaling analysis and confirm that the central charge of the theory takes the value of $c=1/2$, as expected for the Ising universality class with one fermionic degree of freedom. In the explored temperature range, we discover three regimes. A \emph{para} regime, with no local order parameter, a \emph{disordered} regime with local order in imaginary time - corresponding to a local moment formation - , and an antiferro regime, analogous to the domain wall regime of the Ising model\cite{Sachdev1997}. In the third regime, the particles are mostly trapped in one of the minima and form domains due to thermal fluctuations. Since this is a one dimensional model, at sufficiently large system sizes and infinite simulation times, the domains would on average restore the symmetry, and the order parameter would vanish on both sides of the QCP at any finite temperature.

This demonstrates that the R\'enyi entropy indeed detects the local moment formation. We find that the line separating the para and disordered regimes, as identified by the maximum of R\'enyi entropy, exhibits no temperature dependence. This behavior is analogous to the spin-freezing crossover, described by Werner \emph{et al.}\cite{Millis2008} for fermionic systems, and detectable through imaginary time correlations. On the contrary, the crossover between the disordered and antiferro regimes shows a temperature dependence, with classical fluctuations aiding to break the global symmetry at finite system sizes. 

The rest of the paper is structured as follows. In Sec.~\ref{sec:Renyi entropy}, we describe the new implementation of the algorithm for the evaluation of R\'enyi entropy at finite temperature with Langevin dynamics. In Sec.~\ref{sec:Chain of anharmonic oscillators}, we consider the coupled anharmonic chain and evaluate R\'enyi entropy across the phase transition. By studying the subsystem-size scaling, we conclude that the transition belongs to the two-dimensional (2D) Ising universality class. Furthermore, we show the importance of extrapolating the results to the zero imaginary time step, otherwise affected by a sizable imaginary time discretization bias. Finally, in Sec.~\ref{sec:Crossover phase diagram at low temperature} we explore the non-zero temperature phase diagram and discuss the crossover properties between the three regimes. In Sec.~\ref{sec:Conclusion}, we present our conclusions and draw the connections between the described regimes and some phases observed in real systems such as water ice. Some further discussions, technical details, and proofs are left in the Appendices.

\section{R\'enyi entropy and path integrals}
\label{sec:Renyi entropy}

Quantum R\'enyi entropy is a generalisation of quantum Von Neumann entropy, with free parameter $\alpha\in\mathbb{R}_{>0}\setminus\{1\}$. For a bipartite system $A\cup B$, the definition of the R\'enyi entropy of subsystem $A$ and order $\alpha$, $S_A^{\alpha}$, is based on the the reduced density matrix $\hat{\rho}_A$ of the partition $A$ ($\hat{\rho}_A=\text{Tr}_B\rho$) raised to the power $\alpha$ and traced out over the $A$ degrees of freedom. More precisely,
\begin{equation}
S_A^{\alpha} = \frac{1}{1-\alpha}\log\frac{\text{Tr}_A\big[\hat{\rho}_A\big]^{\alpha}}{ [\text{Tr}\hat{\rho}]^{\alpha}},
\label{eq:RenyiEntropyDef}
\end{equation}
where $\text{Tr}\hat{\rho}$ is the normalization constant, ensuring that the the density matrix has unit trace. In the limit of $\alpha\to1$ the entropy equals the Von Neumann entropy. The R\'enyi entropy fulfills almost all essential properties of entropy, except for some inequalities, like the subadditivity and triangle inequality \cite{Linden2013}. This does not affect its performance as an entanglement measure in a pure state \cite{Song2016}. The reason for the R\'enyi entropy popularity is its simplicity. In particular, the evaluation of the collision entropy ($\alpha=2$), defined as a logarithm of the purity $\text{Tr}[\hat{\rho}^2]$, is considerably simpler if compared to the evaluation of $\text{Tr}[\hat{\rho}_A\log\hat{\rho}_A]$.

By considering $\alpha$ replicas of the system, R\'enyi entropy can be recast as the average of a swap operator (when $\alpha=2$) or a circular permutation operator (for $\alpha>2$), acting on the subsystem of interest. When expressed in path integral formalism, these operators reduce to the free energy of merging imaginary-time trajectories (\textit{world-lines}) for the subsystem of interest. The remaining difficulty of evaluating the logarithm of observables that can still vary over several orders of magnitude has been addressed by recent works\cite{Humeniuk2012,Alba2017,Demidio2020,DEmidio2022,Luitz2014,Buividovich2008,Zhao2022new,Zhao2022, Srdinsek2021}. In one of them, we proposed a thermodynamic integration scheme based on regularizing paths which  significantly reduces the R\'enyi entropy variance, yielding low-error low-bias averages. In the following (Sec.~\ref{subsec:Path integral}), we first recall definitions and how expressing the density matrix in terms of path integrals can be used to map the R\'enyi entropy evaluation problem to the one of merging ring polymers. Then, in Sec.~\ref{subsec:Path Integral Ornstein-Uhlenbeck Dynamics (PIOUD)} we present a new algorithm that allows the inclusion of the path-regularization scheme introduced by us in Ref.\cite{Srdinsek2021} into the framework of path integral Langevin dynamics, used to sample the quantum thermal distribution.

\subsection{Path integrals}
\label{subsec:Path integral}

In the path integral formulation, the quantum Hamiltonian of $N$ degrees of freedom of the form
\begin{equation}
\label{eq:NparticleHamiltonian}
\hat{H} = \sum_i^N \frac{\hat{p}_i^2}{2m_i}+\hat{V}(q_1, q_2, ..., q_N),
\end{equation}
where $q_i$ and $p_i$ are positions and momenta of the $i$-th particle with mass $m_i$, is mapped to an analogous classical model. The resulting classical canonical partition function is expressed as
\begin{equation}
\label{eq:DensityMatrixClassicalIsomorphism}
\textbf{Tr}\big[e^{-\beta \hat{H}}\big] = \Big(\frac{1}{2\pi\hbar}\Big)^P\int d^f\mathbf{p}d^f\mathbf{q}e^{-\zeta H_P(\mathbf{p}, \mathbf{q})},
\end{equation}
where $f=PN$, and $P$ the number of \emph{beads} forming rings, each one representing a quantum particle. The accuracy of the mapping grows with the inverse of $\zeta=\beta/P$, the imaginary time step separating neighboring beads, with a convergence rate dependent on the observable. The analogous classical Hamiltonian is given by
\begin{eqnarray}
\label{eq:HamiltonianClassicalIsomorphism}
H_P(\mathbf{p}, \mathbf{q}) =\nonumber\\ \sum\displaylimits_{i=1}^{N}\sum\displaylimits_{j=1}^{P}\Big(\frac{(2\pi\hbar)^2}{2m_i}[p_i^{(j)}]^2 + \frac{1}{2}m_i\omega_P^2(q_i^{(j)}-q_i^{(j+1)})^2\Big) \nonumber\\+ \sum\displaylimits_{j=1}^{P}V(q_1^{(j)}, ..., q_N^{(j)}),
\end{eqnarray}
and $\omega_P = 1/\zeta\hbar$. The upper index (imaginary time slices) has periodic boundary conditions, with period $P$ ($q_i^{(P+1)}=q_i^{(1)}$). In this way, the particles are mapped to harmonic rings, where each bead belongs to an imaginary time slice. Inter-particle interaction involves only beads with the same upper index. 

Looking back at Eq.~\eqref{eq:RenyiEntropyDef}, we see that the denominator describes an ensemble with $\alpha$ rings of length $P$ for each particle. Also the trace over $B$ in the numerator results in the particles belonging to $B$ to form rings of length $P$. However, the multiplication of the reduced density matrix $\hat{\rho}_A$ with itself results in the ensemble of $|A|$ rings of length $\alpha P$. The effect of the traces is therefore visible in the imaginary time boundary conditions. 

Hence, the free energy cost of changing boundary conditions equals the R\'enyi entropy. The evaluation of free energy differences is a very common computational problem for which many algorithms exist. It is therefore no surprise that a whole arsenal of these methods was already applied to the quantum R\'enyi entropy evaluation in Monte Carlo simulations \cite{Zhao2022new,Zhao2022,Srdinsek2021,Demidio2020,Alba2017, Luitz2014,Herdman2014a,Kallin2013,Inglis2013,Alba_2013,Humeniuk2012,Singh2011,Hastings2010,Buividovich2008}. One of the methods, particularly suitable for large or continuous systems, is the recent extension of the thermodynamic integration, where regularised paths are used. In this method \cite{Srdinsek2021}, one expresses the entropy as an integral over $\lambda$ in such a way that
\begin{eqnarray}
\label{eq:Integration}
\log\frac{\text{Tr}\rho_A^{\alpha}}{ [\text{Tr}\rho]^{\alpha}}=\nonumber\\\int\displaylimits_0^1\partial_{\lambda}\log\Big(\mathcal{Z}[\lambda]\Big)d\lambda= -\zeta\int\displaylimits_{0}^{1}\Big\langle \partial_{\lambda}H(\lambda)\Big\rangle_{\mathcal{Z}[\lambda]} d\lambda,
\end{eqnarray}
where the boundary conditions in imaginary time of the Hamiltonian of the system are smoothly deformed so that $\mathcal{Z}[0]=[\text{Tr}\rho]^{\alpha}$ and $\mathcal{Z}[1]=\text{Tr}\rho_A^{\alpha}$. The entropy is then defined as the work required to change the boundary conditions from $\alpha$ replicas of independent rings to the system where each particle belonging to system $A$ is merged into a single ring. In the simulation, the $\lambda$-integral is performed numerically on a finite grid. However, the path has to be chosen wisely in order to result in a low-variance estimate.

We use the same path as the one described in Ref.~\cite{Srdinsek2021}, where only the interactions that enforce the boundary conditions are varied. The classical Hamiltonian along the path is written as
\begin{eqnarray}
\label{eq:PathHamiltonian}
    &H_i(\lambda)=H_{\alpha P}(\mathbf{p}, \mathbf{q})\nonumber \\&+ \sum_{i\in A}\sum_{j=1}^{\alpha}g(\lambda)(\textbf{q}^{(jP)}_{i}-\textbf{q}^{(jP-P+1)}_{i})^2 \nonumber\\&+(h(\lambda)-1)(\textbf{q}^{(jP)}_{i}-\textbf{q}^{(jP+1)}_{i})^2 ,
\end{eqnarray}
where $H_{\alpha P}$ is the extended Hamiltonian of fused rings given by Eq.~\eqref{eq:HamiltonianClassicalIsomorphism} for the particles belonging to the subsystem $A$, while the particles in $B$ keep the boundary conditions of $\alpha$ disconnected ensembles. For particles in $A$, the upper index is periodic with period $\alpha P$ ($q_i^{(\alpha P+1)}=q_i^{(1)}$). $H(\lambda)$ lives on a plane spanned by the coordinates $g$ and $h$, with the path described by the prescription
\begin{equation}
\label{eq:Path}
    (g, h) = ((1-\lambda)^3, \lambda^3).
\end{equation}
In Sec.~\ref{subsec:Path Integral Ornstein-Uhlenbeck Dynamics (PIOUD)} we present a Langevin dynamics algorithm that is able to simulate the ensembles along the presented integration path. 

\subsection{Path Integral Ornstein-Uhlenbeck Dynamics (PIOUD)}
\label{subsec:Path Integral Ornstein-Uhlenbeck Dynamics (PIOUD)}

Path integral molecular dynamics (PIMD) algorithms aim at sampling ring polymer configurations through dynamical trajectories\cite{Marx1996,markland_nuclear_2018}. Among those, Path Integral Ornstein-Uhlenbeck Dynamics (PIOUD) uses dynamical equations\cite{Mouhat2017} driven by the forces acting on the coordinates along the trajectory, and a Langevin thermostat. Langevin dynamics equations are solved in order to introduce the coupling with a thermal bath. This approach is free from the need of designing smart Monte Carlo moves and is therefore much more transferable to the simulations of more complex higher dimensional models. Nevertheless, solving the Langevin equations of motion should be done efficiently.

Sampling the phase space according to the canonical distribution is achieved by evolving the following stochastic equations of motion:
\begin{align}
\label{eq:StochasticDifferentialEquation}
\dot{\tilde{\mathbf{p}}}&=-\gamma \tilde{\mathbf{p}} - \mathbf{\underline{K}} \tilde{\mathbf{q}} + \mathbf{f}(\tilde{\mathbf{q}}) + \boldsymbol{\eta}(t)\\
\dot{\tilde{\mathbf{q}}}&=\tilde{\mathbf{p}}\nonumber,
\end{align}
where $\tilde{\mathbf{p}}$ and $\tilde{\mathbf{q}}$ are $DNP\alpha$-dimensional vectors of momenta and positions of the whole system rescaled by mass ($\tilde{p}_i = p_i/\sqrt{m_i}$ and $\tilde{q}_i = q_i\sqrt{m_i}$), and $\textbf{f}=-\nabla_{\tilde{\mathbf{q}}}V$ is the vector of forces obtained from the Born-Oppenheimer (BO) potential energy surface $V$. The only stochastic contribution comes from $\boldsymbol{\eta}$, a normally distributed noise with zero mean and unit variance. Together with the damping, controlled by $\gamma$, these two terms fix the temperature in accordance with the fluctuation-dissipation theorem. Usually, $\gamma$ is a diagonal matrix. In a more general setting, where forces are sampled stochastically because they are affected by intrinsic errors, such as the ones evaluated, for instance, by electronic quantum Monte Carlo methods\cite{Sorella2008}, it can also have off-diagonal terms, related to the force covariance matrix.

\begin{figure}[t]
\centering
\includegraphics[width=0.9\linewidth]{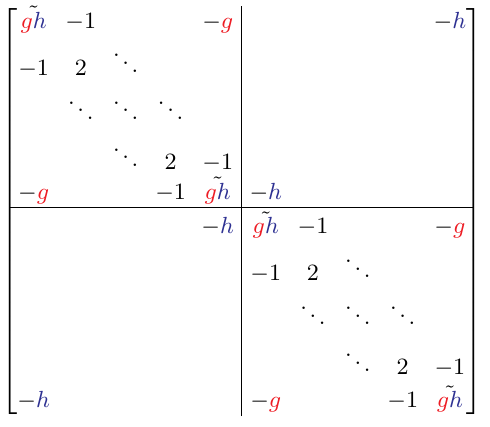}
\caption{The matrix $\Tilde{K}(\rsplit{g},\rjoint{h})$ for $\alpha=2$. Abbreviation $\tilde{\rsplit{g}\rjoint{h}}=1 + \rsplit{g} + \rjoint{h}$ is used.\label{fig:Matrix}}
\end{figure}

On the other hand, $\mathbf{\underline{K}}$ is an off-diagonal matrix in the imaginary time indices, which represents the harmonic interaction in imaginary time. In the normal path integral ensemble driven by the Hamiltonian in Eq.~\eqref{eq:HamiltonianClassicalIsomorphism}, it is simply written as
\begin{equation}
\label{eq:KmatrixOriginal}
K_{il}^{(j)(k)}=\omega_P^2\delta_{il}\Big(2\delta^{(j)(k)}-\delta^{(j)(k-1)}-\delta^{(j)(k+1)}\Big).
\end{equation}
Yet, in our case the matrix depends on the lower index as well. Indeed, particles belonging to subsystem $A$ have modified boundary conditions, that are functions of the parameters $g$ and $h$ [Eq.~\eqref{eq:Path}]. The new $\mathbf{\underline{K}}$ matrix should therefore be
\begin{equation}
\label{eq:KmatrixModified}
\omega_P^2\delta_{il}\Big(\delta^{i\in A}\Tilde{K}^{(j)(k)}(g, h)+\delta^{i\notin A}\Tilde{K}^{(j)(k)}(1, 0)\Big),
\end{equation}
when subsystem $A$ is considered in the evaluation of the R\'enyi entropy. The upper index is again periodic, with period $\alpha P$.  The new smaller matrix $\Tilde{K}^{(j)(k)}(g, h)$ includes all the ensembles described by Eq.~\eqref{eq:PathHamiltonian}. It is a function of parameters $g$ and $h$, defined as
\begin{widetext}
\begin{equation}
\label{eq:KmatrixModifiedDetailed}
\Tilde{K}^{(j)(k)}(g, h)=\left\{\begin{array}{l} \omega_P^2\Big((1 + g + h)\delta^{(j)(k)}-h\delta^{(j)(k-1)}-\delta^{(j)(k+1)} -g\delta^{(j)(k+P-1)}\Big) ~~~ \textrm{for $(j\mod P) = 0$}\\
\omega_P^2\Big((1 + g + h)\delta^{(j)(k)}-\delta^{(j)(k-1)}-h\delta^{(j)(k+1)} -g\delta^{(j)(k-P+1)}\Big) ~~~ \textrm{for $(j\mod P) = 1$}\\
\omega_P^2\Big(2\delta^{(j)(k)}-\delta^{(j)(k-1)}-\delta^{(j)(k+1)}\Big) ~~~ \textrm{otherwise}.
\end{array}\right.
\end{equation}
\end{widetext}
The structure of the $\Tilde{K}$ matrix is illustrated in Fig.~\ref{fig:Matrix}.

The forces in Eq.~\eqref{eq:StochasticDifferentialEquation} come from both the harmonic and inter-particle interactions, thus they span very different energy scales. This renders the sampling very inefficient. The solution employed in the PIOUD algorithm is to split the time evolution in two operators. Indeed, the time evolution described by the action of the Fokker-Planck Liouville operator is split with symmetric trotterization to yield
\begin{equation}
\label{eq:Trotterization}
e^{i\mathcal{L}\delta t}=e^{i\mathcal{L}^{BO}\delta t/2}e^{i\mathcal{L}^{harm}\delta t}e^{i\mathcal{L}^{BO}\delta t/2}+\mathcal{O}(\delta t^3),
\end{equation}
where $\mathcal{L}^{BO}$ contains the propagation of the particles interacting at each imaginary time slice through the BO forces $\textbf{f}$, neglecting the harmonic interaction between the beads, which is instead included in $\mathcal{L}^{harm}$. Both $\mathcal{L}^{BO}$ and $\mathcal{L}^{harm}$ contain stochastic and dissipation terms, where the $\gamma$ matrix is adjusted for each propagator. While in the BO ensemble the $\gamma$ matrix is used as a user-defined constant ($\gamma^{BO}$), in the harmonic one it is chosen according to the optimal damping scheme for harmonic oscillators\cite{Rossi2014}. This corresponds to writing it in the eigenbasis of the matrix $K$ in Eq.~\eqref{eq:KmatrixOriginal}, where $\gamma$ is a diagonal matrix with elements\cite{Mouhat2017}
\begin{equation}
\label{eq:GammaMatrixElements}
\gamma_{harm}^{(k)}=\left\{\begin{array}{ll}2\Omega_k & \text{if } 2\Omega_k\geq\gamma_0,\\
\gamma_0 & \text{otherwise.}\end{array}\right.
\end{equation}
$\Omega_k$ are eigenvalues of matrix $K$, and now the upper index $(k)$ indicates the corresponding $k$-th eigenvector in the beads space. This choice for $\gamma_{harm}$ not only optimizes the dumping process but also guarantees that $[\gamma_{harm},K]=0$. This latter condition is needed to integrate exactly the Ornstein Uhlenbeck dynamics, i.e. the thermalized Brownian quantum diffusion, encoded in $\mathcal{L}^{harm}$\cite{Mouhat2017}. As in the BO ensemble, the damping coefficient related to the center of mass dynamics ($\gamma_0$) has to be found by optimizing the diffusion coefficient of the process by running short simulations. Once the optimal value found, it is transferable to similar systems. 

However, with the new $\mathbf{\underline{K}}$ matrix of Eq.~\eqref{eq:KmatrixModified}, different particles have different eigenvalues and eigenvectors, depending on which subsystem they belong to. The same should therefore hold also for the $\gamma$ matrix, which now depends on the particle (lower) index as well, since the $\mathbf{\underline{K}}$ eigenvalues depend on it. Indeed, what we propose is to simply extend the prescription in Eq.~\eqref{eq:GammaMatrixElements}, and write 
\begin{equation}
\label{eq:GammaMatrixElementModified}
[\gamma_{harm}^{(k)}]_i=\left\{\begin{array}{ll}2\Omega^{(k)}_i & \text{if } 2\Omega^{(k)}_i\geq\gamma_0,\\
\gamma_0 & \text{otherwise.}\end{array}\right.
\end{equation}
This has profound implications, as it shows that this type of Langevin thermalization cannot work in an integration scheme where the propagation of the harmonic part is not diagonal in the lower index. Nevertheless, our choice of $\gamma_{harm}$ in Eq.~\eqref{eq:GammaMatrixElementModified} still fulfills the condition $[\gamma_{harm},K]=0$ by construction. Therefore, $\mathcal{L}^{harm}$ can be exactly integrated even with the optimal dumping scheme of Eq.~\eqref{eq:GammaMatrixElementModified}, appropriately generalized for the extended Hamiltonian in Eq.~\eqref{eq:PathHamiltonian}.

The algorithm then proceeds as follows. First, the eigenvalues of $\Tilde{K}^{(j)(k)}(g, h)$ are found for either of the two cases in Eq.~\eqref{eq:KmatrixModified}. The propagator $e^{i\mathcal{L}^{harm}\delta t}$ can be evaluated exactly if written in the $\mathbf{\underline{K}}$ eigenbasis, owing to the fact that there are no contributions from physical forces $\textbf{f}$ present. From here on the algorithm consists of rotating the coordinates back and forth between the eigenbasis of $\Tilde{K}^{(j)(k)}(g, h)$ and the coordinate basis, according to the following prescription:
\begin{enumerate}
    \item Update the particles momenta by applying the $e^{i\mathcal{L}^{BO}\delta t/2}$ propagator, according to the equation
    \begin{eqnarray}
       \label{eq:PropagationOfMomentum}
        \tilde{\mathbf{p}}(t) = e^{-\gamma^{BO}\delta t/2}\tilde{\mathbf{p}}(t-\delta t/2)\nonumber\\+\int\displaylimits_{t-\delta t/2}^tdt'e^{\gamma^{BO}(t'-t)}[\textbf{f}(t-\delta t/2) + \boldsymbol{\eta}(t')];
    \end{eqnarray}
    \item Transform the vectors of positions and momenta of each particle from the coordinates basis to the $\Tilde{K}^{(j)(k)}(g, h)$ eigenbasis;
    \item Propagate them exactly, by means of the $e^{i\mathcal{L}^{harm}\delta t}$ propagator;
    \item Perform the inverse transformation, back to the coordinate basis;
    \item Evaluate the forces coming from the physical potential as $\textbf{f}(t+\delta t)=-\nabla_{\tilde{\mathbf{q}}}V(t+\delta t)$,
    \item Close the symmetric form by applying the $e^{i\mathcal{L}^{BO}\delta t/2}$ propagator in Eq.~\eqref{eq:PropagationOfMomentum} again, according to Eq.~\eqref{eq:Trotterization}.
\end{enumerate}

Further numerical details are given in Appendix~\ref{appendix:Numerical details}. Also, as molecular dynamics (MD) is not always efficient to sample phase space when large energy barriers are present, we have introduced a SWAP move (see Appendix~\ref{appendixA:Accelerating MD sampling with SWAP operator}), performed randomly. Thus, the sampling performed is a hybrid MD and Monte-Carlo scheme. The algorithm was tested on integrable entangled model of two coupled harmonic oscillators (see Appendix~\ref{appendixB:Tests of the method}).

\section{Chain of anharmonic oscillators: Zero temperature phase diagram}
\label{sec:Chain of anharmonic oscillators}

\subsection{System description}

The discrete $\phi^4$ model, where the field $\phi$ is discretised in space, analogous to an ultraviolet cutoff, is described by the Hamiltonian in Eq.~\eqref{eq:NparticleHamiltonian}, with the potential term
\begin{equation}
\label{eq:DoubleWellPotential}
\hat{V}(\hat{\textbf{q}}) = \sum_i^N \frac{\theta}{2}(\hat{q}_i-\hat{q}_{i+1})^2-m\omega^2\hat{q}_i^2+\lambda\hat{q}_i^4,
\end{equation}
where $q_i=\phi(i)$ are values of the field at fixed positions $i$. It depicts a chain of $N$ particles trapped in an external double-well potential of the form $U = (\theta-m\omega^2)\hat{q}_i^2 + \lambda \hat{q}_i^4$. The coupling between the particles comes from the harmonic interaction ($-\theta \hat{q}_i\hat{q}_{i+1}$). In this work, we consider a particular limit of the discrete $\phi^4$ model, where the quadratic term is removed, resulting in a single-well anharmonic potential (Appendix~\ref{appendix:phi4 model}). This is achieved by fixing $\theta = m\omega^2$ and rescaling the mass through $m\to m\lambda$, thus reducing the number of free parameters to one, such that
\begin{equation}
    \label{eq:ChainOfDoubleWellsPotentialMinus}
    \hat{V}(\hat{\textbf{q}})=\sum_i^N \hat{q}_i^4 - D^2\hat{q}_i\hat{q}_{i-1},
\end{equation}
in the system with periodic boundary conditions ($\hat{q}_{N+1}=\hat{q}_{1}$), with $D^2=m\omega^2$. In fact, $D^2$ can be either negative (the "ferro" case) [Eq.~\eqref{eq:ChainOfDoubleWellsPotentialMinus}] or positive (the "antiferro" case):
\begin{equation}
    \label{eq:ChainOfDoubleWellsPotential}
    \hat{V}(\hat{\textbf{q}})=\sum_i^N \hat{q}_i^4 + D^2\hat{q}_i\hat{q}_{i-1},
\end{equation}
where the negative sign in front of $D^2$ can be interpreted as coming from a potential term with negative curvature (i.e. imaginary phonon) at $q=0$. Hereafter, we will adopt the latter situation [Eq.~\eqref{eq:ChainOfDoubleWellsPotential}], as reference Hamiltonian for the chain of anharmonic oscillators. 

The system was shown to undergo a quantum phase transition of the order-disorder type in the double well regime, and of the displacive type in the anharmonic regime. In both cases the transition belongs to the 2D Ising universality class\cite{Milchev1986,Toral1990,Wang1994,Rubtsov2001,Savkin2002,Kim2007,Barsan2008}. In fact, in the limit of very deep wells, it describes a two level system, analogous to the quantum Ising model in a transverse field. 

The model is extensively studied also in the continuous limit ($d/N\to0$, $d$ being the lattice spacing), describing a relativistic quantum scalar field, where the interaction term in Eq.~\eqref{eq:DoubleWellPotential} reduces to $(\partial_x q(x))^2$. It is arguably the simplest model that contains kinks, defined as abrupt changes of the field configuration jumping from one minimum of the double well to the other. In this limit the renormalisation of the diverging quadratic term reduces the number of free parameters to one\cite{Harindranath1988,Lee2001,Sugihara_2004,Schaich2009,Milsted2013,Rychko2015}. Phase transition can be observed also in the continuous case. However, the critical value of the parameter is not agreed upon\cite{Rychko2015}. 

\begin{figure*}[t]
\centering
\includegraphics[width=\linewidth]{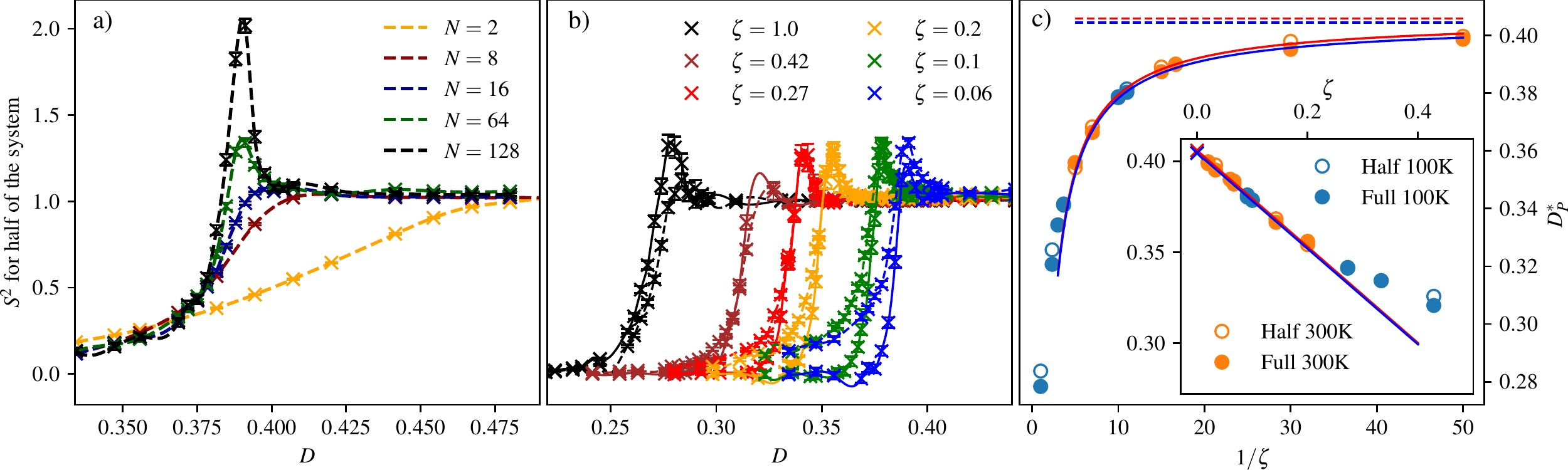}
\caption{R\'enyi entropy as a function of $D$. a) R\'enyi entropy of half of the system for different system sizes $N$ at $\zeta=0.06$. Peak forming at the phase transition point appears when $N$ is sufficiently large. b) Critical $D^*$ depends on how many beads we use. This is demonstrated at $100$ K for the entropy of the full system (\textit{solid lines}) and half of the system (\textit{dashed lines}). Also the level of quantum correlations, $S^2_{half}-S^2_{full}$, is affected. c) $D^*_P$ dependence on the reciprocal of ratio $\zeta = 300 / PT$. The plot includes points extracted from data at $100$ K (\textit{blue color}) and $300$ K (\textit{orange color}), which seem to agree, for the peaks of the entropy of half of the system (\textit{empty circles}) and the full (\textit{filled circles}) system. In the inset, a clear dependence on $\zeta$ is demonstrated. By a linear extrapolation in $\zeta$, the critical $D$ is predicted to be $D^*_{\infty}=0.405\pm0.001$. Continuous curves in a-b) are splines fitted on the data. They have been used to determine the maxima. \label{fig:EntropyScan}}
\end{figure*}

The discrete chain is particularly interesting because it approximates many condensed matter systems, depending on the interpretation of coordinates $q_i$. If interpreted as displacements of protons in the system, the model can describe hydrogen halides, like HF and HBr\cite{Jansen1987,Springborg1988,Wang1994}, super-hydrides, such as LaH$_{10}$\cite{Drozdov2019}, YH$_n$\cite{Kong2021}, H$_3$S\cite{Drozdov2015}, and even water ice phases, such as VII, VIII, and X\cite{Holzapfel1972,Wendy2002,Pruzan2003,Reinhardt2022}. On the other hand, if $q_i$ are interpreted as local dipoles, the same model can represent quantum dielectric materials\cite{Hotta2022}, such as $\kappa$-ET$_2$Cu$_2$(CN)$_3$ and $\kappa$-ET$_2$Cu[N(CN)$_2$]Cl, possessing electronic ferro-electricity\cite{Lunkenheimer2012}. In magnetic materials, such as LiHoF$_4$\cite{Bitko1996}, the $q_i$ coordinates represent local magnetic moments.

According to the classical analysis, the system features two degenerate global minima, with ferro (antiferro) order if the coupling constant is purely imaginary (real). The two models in Eqs.~\eqref{eq:ChainOfDoubleWellsPotentialMinus} and \eqref{eq:ChainOfDoubleWellsPotential} are equivalent up to the symmetry transformation of flipping every second coordinate. As already mentioned, here we choose the antiferroelectric situation for illustrative purpose in Sec.~\ref{sec:Crossover phase diagram at low temperature}. The following solution represents the two minima in the antiferro case:
\begin{equation}
\label{eq:ChainOfDoubleWelsMinima}
    q^m_{2i}=\pm\frac{D}{\sqrt{2}},\;\;q^m_{2i-1}=\mp q^m_{2i},
\end{equation}
if the number of oscillators $N$ is even. A system with odd number of oscillators is frustrated and avoided in our analysis. When $N>8$, the classical system can move from one minimum to the other through the creation of kink-antiking pair, by flipping only one particle. Once a kink-antikink pair is created, it can grow a domain without additional energy cost. The classical energy cost of kink-antikink pair creation is $\Delta V=2 D^4$, as opposed to the collective crossing of the saddle point, when the cost increases with the system size $N$ as $\Delta U=N D^4 / 4$. By increasing $D$, particles are pushed further away, while the height of the barrier increases. In the following, we study the system at temperatures much lower than the height of the barrier, such that $k_bT\ll\Delta V$. Thus, by choosing $\theta = m\omega^2$ in Eq.~\eqref{eq:DoubleWellPotential}, we have one external parameter, $D$, that tune the height of the potential energy barrier, similar to pressure in, say, ice. Also, with this choice, there is no phase transition at $T=0$ in the classical limit, as the barrier disappears only at $D=0$. However, due to quantum fluctuations, the particles can tunnel through the barrier for some values of $D$, and thus on average restore the symmetry $\langle q\rangle=0$. With the new algorithm introduced in Sec.~\ref{sec:Renyi entropy}, we aim at pin-pointing the position of the QCP using the R\'enyi entropy of the system. We will show that the $D$ dependence of the entropy strongly resembles the one of the Ising model criticality and can be used to locate the phase transition, and evaluate its central charge. In order to simplify the comparison with realistic systems, we fix the energy scale by considering oscillators with the proton mass $m_p=1837.1799$ in atomic units and the potential given in Eq.~\eqref{eq:ChainOfDoubleWellsPotential}. Indeed, for such choice we observe a phase transition at physical H-H distances and physical height of the potential barrier of 15 kJ/mol$^{-1}$ per hydrogen bond.

\subsection{Quantum critical point}

As argued in the Introduction, the entropy of a subsystem at zero temperature in 1D grows with the subsystem size only up to a certain threshold. This rule is violated only in the vicinity of the QCP, where the growth is logarithmic and never saturates\cite{Eisert2010}. One can also reverse this point and claim that by looking at the entropy of a fixed subsystem size, where it is sufficiently large (for example half of the system), one could spot the critical point by the spike in the entropy. Upon increasing the system size, one should see that the entropy saturates everywhere except in the vicinity of the phase transition. Given that the critical system is gapless, the entropy should diverge exactly at the QCP. Extracting the peak is therefore often sufficient for its detection\cite{Irenee2016,Sharma2022,Latorre2005,Lambert2004,Serwatka2022}. However care has to be taken, to show that away from the peak the entropy eventually saturates. This will be done, after we fully resolve the position of the peak as a function of the system size $N$ and imaginary-time discretization $\zeta$. The same can be said also about the entropy of the system at finite (sufficiently low) temperature. A very nice example of this is the 1D Ising model, where the peak of the entropy at finite temperature exactly coincides with the QCP, at relatively small system sizes\cite{Srdinsek2021}.

In Fig.~\ref{fig:EntropyScan}a we scan over the parameter $D$ at $100$ K and plot the entropy of half of the system. By increasing the system size, the appearance of the peak is clearly visible, with the thermodynamic limit reached already at $N=64$. The peak is positioned exactly where it would be expected, at the point where an abrupt change in entropy appears. Because the temperature is not exactly zero, the entropy computed here does not quantify only pure quantum fluctuations.

So far, we neglected the effect of the discretisation in the imaginary time. The phase transition to a para phase is driven by quantum fluctuations, which restore the symmetry of the ground state, while the discretisation in imaginary time exactly truncates these fluctuations. Thus, at finite $P$ the restoration of the symmetric para state will occur slower than in the $P\rightarrow\infty$ limit, and the critical value of $D$, dubbed $D^*_P$, will be always smaller than $D^* \equiv D^*_\infty$. This can be seen like performing semi-classical approximations by introducing a finite $P$ instead of performing an expansion in the powers of $\hbar$.

\begin{figure*}[t]
\centering
\includegraphics[width=\linewidth]{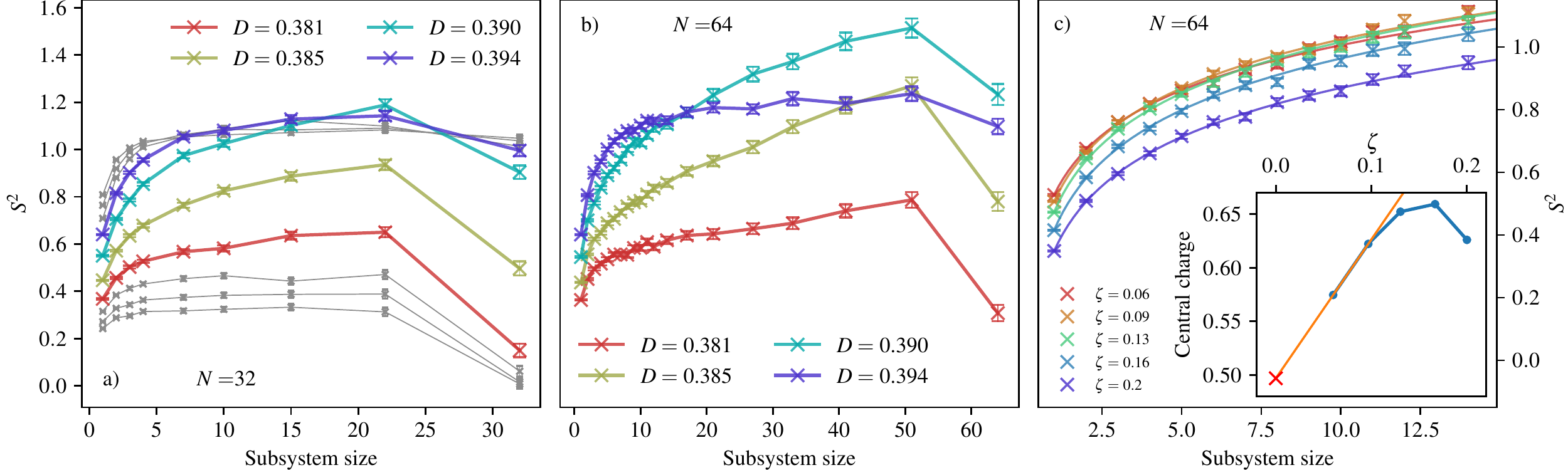}
\caption{R\'enyi entropy as a function of subsystem size in the neighborhood of the phase transition at 100 K. a-b) Scaling for the $N=32$ (a) and $N=64$ (b) chains for equally spaced values of $D$, with solid lines connecting the points. The colored sets of points are in the vicinity of the phase transition. It is clearly visible that for a general $D$ the entropy saturates, while it keeps growing in the critical regime. c) Scaling for the $N=64$ chain at critical $D$, for various values of the ratio $\zeta = 300 / PT$. The solid lines in the panel c) are fits of Eq.~\eqref{eq:ScalingOfEntropy} to the points. Effects are relatively small and convergence is fast. In the inset, we show the parameter in front of the fitted logarithmic terms as a function of $\zeta$. By a crude linear extrapolation, $c_{\zeta=0}$ is extracted.
\label{fig:EntropyScaling}}
\end{figure*}

The discretisation error typically depends on the parameter $\zeta=\beta / P$, and should be studied for each observable separately\cite{Samson_2000, BRUCH1973143}. Convergence can be slow, or even not reachable for some observables\cite{Hoffman1990}. In our case the effects are quite strong on the position of the phase transition (Fig.~\ref{fig:EntropyScan}b), if compared with the case of $N=2$, where $\zeta=0.06$ already coincides with the analytical result. It can be clearly seen that $\zeta$ shifts the position of the critical point, but also affects the scaling with the subsystem size. By decreasing $\zeta$, the full entropy decreases, which means that the system is more entangled. This is consistent with a better description of the quantumness of the particles.

To extrapolate our results to the exact limit, we studied in Fig.~\ref{fig:EntropyScan}c the dependence of the critical value on $\zeta$. We compared the results at two different temperatures and for the entropy of the half and full system at fixed $N$. By doing so, we can see that the peaks at different temperatures and different subsystem sizes fall on a universal curve (see Fig.~\ref{fig:EntropyScan}c). Moreover, for sufficiently small $\zeta$ the dependence becomes linear in $\zeta$, as shown in the inset of Fig.~\ref{fig:EntropyScan}c. This is not surprising, and it is observed for many convergent observables \cite{Samson_2000, BRUCH1973143}. A linear extrapolation can thus be used to extract the critical value $D^*$ in the limit of $\zeta=0$, which we compare with a previous estimate in Tab.~\ref{tab:table1}. The resulting distance well agrees with the distances observed in realistic systems with strong Hydrogen bond, for which we have previously shown entanglement at room temperature\cite{Pruzan2003,Srdinsek2021}.

\subsection{Scaling of the entropy}

\begin{table}[t]
\caption{\label{tab:table1}%
Values of critical $D$.}
\begin{ruledtabular}
\begin{tabular}{lcdr}
& 
\textrm{\textbf{$\zeta=0$}}&
\multicolumn{1}{c}{\textbf{CIM\footnote{
Cumulant intersection method\cite{Binder1986} at fixed $\zeta$.}\cite{Wang1994}}}\\
\colrule
$D^*$ & $0.405\pm0.001$			& 0.4\\
\end{tabular}
\end{ruledtabular}
\end{table}

The previous procedure could raise some objections, because we analyse the entanglement of a ground state still dressed by thermal fluctuations, although at a very small temperature, as dictated by the use of path integrals. In view of this, we would like to see if the critical point, discovered by inspecting the peak, corresponds to a logarithmic scaling as a function of the subsystem size. In Fig..~\ref{fig:EntropyScaling}a-b we can see that for the values below and above the critical point, the entropy indeed saturates upon increasing the subsystem size. This feature persists even upon increasing the number of particles in the system. 

However, detecting a transition point by looking at the subsystem size scaling would be very time-consuming. One has to blindly extract many subsystem sizes at many couplings. If $N$ is too small, the curves will saturate in an interval of $D$ values that narrows only upon increasing $N$. Additionally, we see that with larger $N$ comes also the need for larger $P$, otherwise the effective temperature increases (Fig.~\ref{fig:EntropyScaling}b). Overall, we can see that the scaling can be used to confirm the correctness of the critical value, but it is not a feasible method for the search of the critical coupling.

Since the entropy clearly diverges close to the QCP and saturates elsewhere, we compared the scaling with the logarithmic curve. Due to the remarkable connection between conformal invariant quantum field theories (CFT) and critical phenomena, the scaling depends on the CFT central charge $c$ of the same universality class\cite{Vidal2003}. The exact dependence is known even for finite temperature and finite system sizes\cite{Calabrese2004}. Written in terms of the R\'enyi entropy of order $\alpha$, the scaling equals\cite{Calabrese2004}
\begin{equation}
\label{eq:ScalingOfEntropy}
S^{\alpha}_l \sim \frac{c}{6}\Big(\frac{\alpha^2-1}{\alpha}\Big)\log(l).
\end{equation}
In Fig.~\ref{fig:EntropyScaling}c, we varied the parameter $\zeta$ and performed the scaling analysis at a $D$ value that corresponds to the peak of the entropy for given $\zeta$. The scaling is clearly in accordance with Eq.~\eqref{eq:ScalingOfEntropy}. From this, the prefactor of the logarithm, directly depending on the central charge, can be extracted. We noticed that, at fixed $\zeta$, the value does not agree with the universality class of the Ising model, $c=0.5$. However, the $\zeta$-dependence analysis suggests that this is again an imaginary-time discretisation effect. Indeed, in the limit of $\zeta=0$, the data agree with the central charge being equal to $0.5$.

\section{Phase diagram at non-zero temperature}
\label{sec:Crossover phase diagram at low temperature}

So far, we have been focusing on the critical phenomena at small fixed temperature with the purpose of locating and characterizing the zero-temperature phase transition. However, the phase diagram around a quantum phase transition usually features rich temperature dependence\cite{Vojta2003,Chakravarty1989,Sachdev1997,Sachdev2011,Gabbrielli2018,Lu2020}. 

Studies of higher dimensional systems found the emergence of classical order-disorder phase transition in the double well chain with finite barrier and a displacive transition at vanishing barrier\cite{Savkin2002}. Investigations in 1D discrete $\phi^4$ models were done in the limit of large double-well barrier, i.e. the Ising model. The transverse Ising model at finite temperature splits in three distinct regimes\cite{Sachdev1997}. For large magnetic field, the ground state is aligned with the magnetic field, while thermal fluctuations flip individual spin in the opposite direction. On the other side of the transition, where the magnetic field is weak, there exists a domain wall regime, with the presence of kink-antikink pairs. Both of these regimes evolve into a quantum critical regime with strong quantum fluctuations. This raises the question whether the same regimes persist after the double well barrier is removed. 

\begin{figure}[b]
\includegraphics[width=\linewidth]{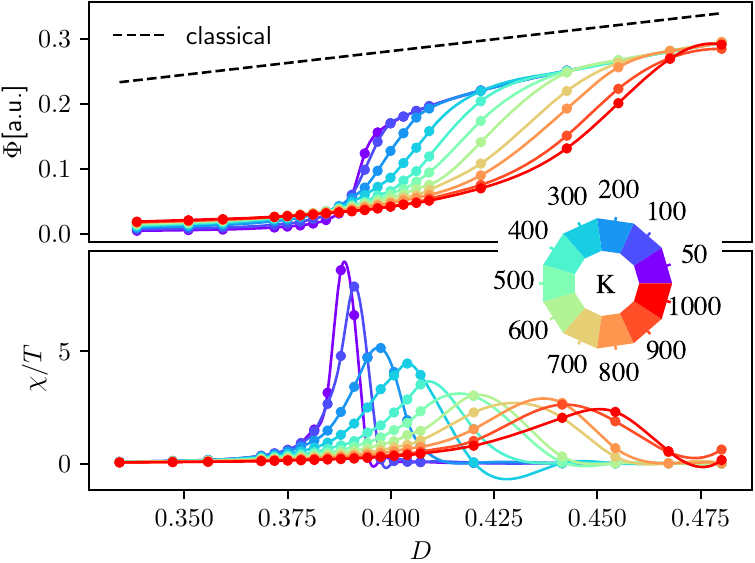}
\caption{Order-disorder order parameter $\Phi$. \textit{Upper panel} - The value of the parameter $\Phi$ across the transition at various temperatures. The restoration of the broken symmetry is marked by a sudden drop of the parameter values. The effect is visible also at higher temperatures. The \textit{dashed} line is the position of the minima of the classical potential. \textit{Lower panel} - Fluctuations (susceptibility) of the order parameter, weighted by the temperature to improve readability. By increasing the temperature the peak of susceptibility function drifts to higher $D$. (\textit{Colors palette:}) from purple at 50 K to red at 1000 K.\label{fig:ClassicalScan}}
\end{figure} 

To address this question, let us now discuss the finite temperature phase diagram of the coupled anharmonic chain, exploiting once more the evaluation of the R\'enyi entropy. Similarly to the Ising chain, we observe three regimes. However, these regimes are different, and we will describe them as para, disordered, and antiferro, as presented in the Introduction. They can be found in this order by raising the coupling $D$. They are separated by two crossovers, that merge in the limit of low temperature at the QCP. Our calculations suggest that the disordered regime disappears close to zero temperature, which results in a direct transition between the antiferro and para regimes.

\subsection{Crossover from antiferro to disordered regime}

Let us start by looking at the strong coupling side of the phase diagram at finite temperature, and progressively reduce $D$. In order to investigate the crossover from antiferro to disordered regime, we consider the total polarization as the order parameter, and extend the study of the ground-state symmetry done by Wang \emph{et al.}\cite{Wang1994} to higher temperatures. For the model in Eq.~\eqref{eq:ChainOfDoubleWellsPotential}, displaying anti-ferromagnetic order, we write the order parameter as
\begin{equation}
\label{eq:Order paramether Gubernatis}
\Phi(D) = \frac{1}{PN} \Big|\sum_{j=1}^P\sum_{i=1}^{N/2}q_{2i}^{(j)}\Big|,
\end{equation}
where the sign of the interaction is taken into account by looking at one of the two bipartite lattices.

Fig.~\ref{fig:ClassicalScan} displays the computed order parameter $\langle\Phi(D)\rangle$ at different temperatures. Two regimes are clearly visible. Due to the presence of quantum fluctuations, the average $\langle\Phi(D)\rangle$ vanishes for small enough $D$ even at the lowest temperature we could access ($50$ K; top panel of Fig.~\ref{fig:ClassicalScan}). The interparticle interaction counteracts the effect of quantum and thermal fluctuations, which have the tendency to restore the symmetry of the potential. By increasing $D$, the value of $\langle\Phi(D)\rangle$ sharply increases and the system undergoes a phase transition (Fig.~\ref{fig:ClassicalScan}). Thermal fluctuations, although smaller than the energy cost of a flip ($k_bT\ll\Delta V$), delay the transition to larger $D$. 

In 1D systems, thermal fluctuations not only delay the transition but also break true long-range order, due to the formation of domain walls (kink-antikink pairs). A way to locate the crossover between the disordered and antiferro regimes is to compute the susceptibility of the order parameter, defined as $\chi=\beta P N \left(\langle \Phi(D)^2\rangle - \langle \Phi(D)\rangle^2\right)$, and look at the position of its maximum. Just as for the R\'enyi entropy, the position of the peak displays a strong dependence on the imaginary time step $\zeta$, with scaling almost identical to the one of the R\'enyi entropy (Appendix~\ref{appendix:The scaling in zeta}). The extrapolated location of the susceptibility peak is shown on the $T-D$ diagram in Fig.~\ref{fig:Phase diagram}. It shows a linear dependence on $D$, even though classical analysis would suggest a dependence on $D^4$. This indicates that its behavior is strongly affected by quantum effects. 

\begin{figure}[t]
\includegraphics[width=\linewidth]{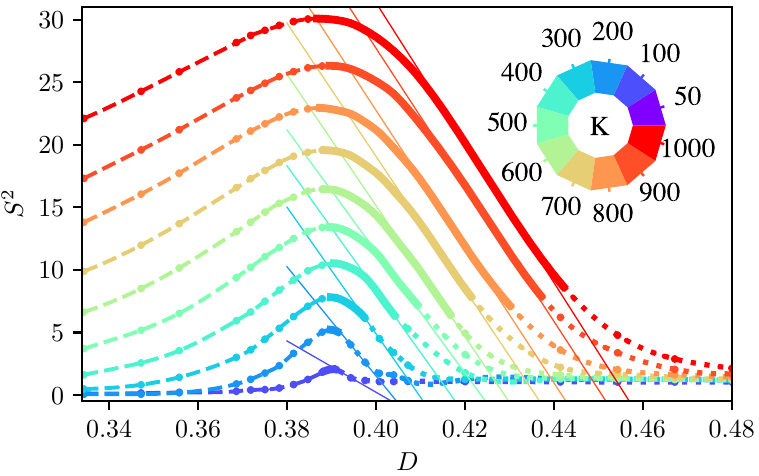}
\caption{Temperature dependence of the R\'enyi entropy. Thermodynamic entropy of a system of $N=128$ particles as a function of coupling $D$ at different temperatures and fixed imaginary time step $\zeta$. Different regimes, described by the order parameter $\Phi$ (Eq.~\eqref{eq:Order paramether Gubernatis}) and imaginary time correlations $C_T$ (Eq.~\eqref{eq:eq:imag_time_corr}), are marked by dashed (para), solid (disordered) and dotted antiferro lines. These regimes are clearly visible in the R\'enyi entropy - Exponential growth of fluctuations with $D$ in para regime, divergence of the R\'enyi entropy at the QCP, almost linear suppression of correlation with increased $D$ in the disordered regime and fast relaxation of entropy to the value of $S^2=1$ in the antiferro regime. (\textit{colors}) From purple at 100 K to red at 1000 K.\label{fig:Renyi_temperature}}
\end{figure} 

We now investigate the full entropy of the system as a function of $D$ and temperature, displayed in Fig.~\ref{fig:Renyi_temperature}. Except for very low temperature, already discussed in Sec.~\ref{sec:Chain of anharmonic oscillators}, we can distinguish three regimes. At low $D$ the R\'enyi entropy is seen to be increasing with $D$, followed by a decrease after reaching a maximum. Then, at even larger values of $D$, the R\'enyi entropy plateaus at $S^2=1$, as expected in the ordered regime, with two degenerate ground states. The crossover between the last two regimes can be estimated as the intersection of a linear extrapolation of the R\'enyi entropy at the inflection point on the right side of the peak, with the $S^2=1$ line. These linear fits are shown in Fig.~\ref{fig:Renyi_temperature} and the crossover values of $D$ at each temperature obtained by this procedure are very close to our estimate of the disorder-to-order crossover from the susceptibility $\chi$.

\subsection{Crossover from disordered to para regime}

The maximum of the R\'enyi entropy in Fig.~\ref{fig:Renyi_temperature} occurs at the strength $D$ corresponding to the critical value for the quantum phase transition observed at low temperature. This suggest that it is the signature of a crossover from a regime where the particles are fully delocalized over the two minima, through a combination of quantum and thermal fluctuations, to a regime where individual particles localize in one basin, with formation of local moments. This is consistent with an expansion of the available phase space, through the formation of a double well, as $D$ increases, and the macroscopic localization in one local minimum as $D$ is increased further. The corresponding increase in the entropy in the first regime and the decrease in the second regime, leads to the maximum at the transition point. Due to thermal fluctuations, this does not result in long range order, but in a regime where the formed local moments are spatially disordered (Fig.~\ref{fig:Phase diagram}). 

\begin{figure}[t]
\includegraphics[width=\linewidth]{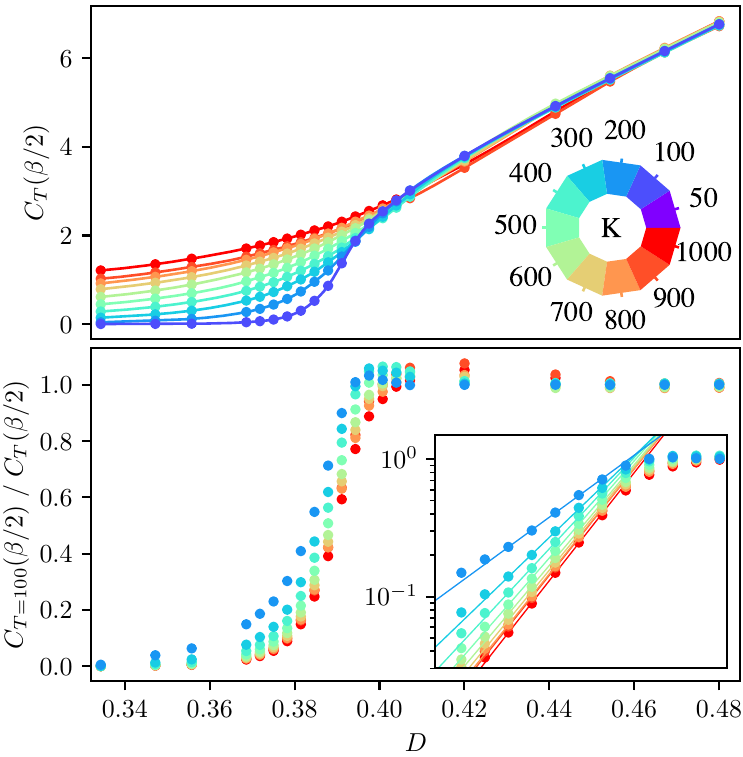}
\caption{Local moment formation. \textit{Upper panel} - The imaginary-time correlation function at half time $C_T(\beta/2)$ for various temperatures. In the local moment regime, $C_T(\beta/2)$ saturates to a finite value. Analogously to the spin freezing\cite{Millis2008}, the temperature dependence of $C_T(\beta/2)$ is suppressed. \textit{Lower panel} - The ratio between $C_T(\beta/2)$ at $T=100$ K and the correlation function at higher temperatures. \textit{Inset} - In logarithmic scale the dependence is linear and the critical $D$ can be estimated. (\textit{Colors palette:}) from dark blue at 100 K to red at 1000 K.\label{fig:Spin-freezing}}
\end{figure} 

\begin{figure}[t]
\includegraphics[width=\linewidth]{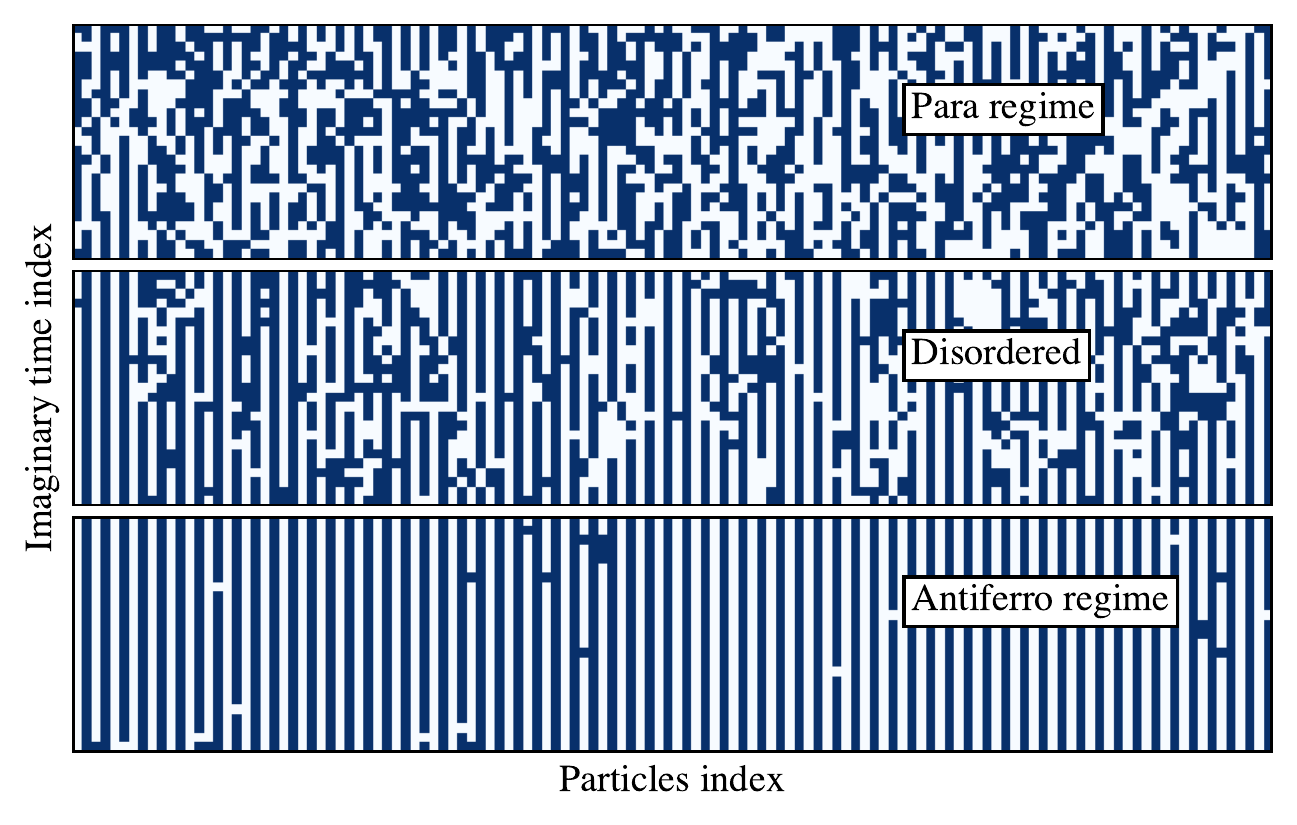}
\caption{Word-lines snapshots in different regimes. Each ring is collapsed according to the sign in front of the bead position to either $-1$ (\textit{dark blue}) or $+1$ (\textit{light blue}). At the point of local moment formation (middle panel), the ordering in imaginary time appears, but it is not frozen. Upon increasing the interaction further, entire rings get trapped to the left- or right-hand side with respect to the central symmetric position. The snapshots are taken at $200 K$ with $25$ beads and $N=128$.\label{fig:Spin-freezing-paths}}
\end{figure}

This crossover thus appears analogous to the spin-freezing phenomenon, described by Werner \emph{et al.}\cite{Millis2008} for strongly correlated fermionic systems, and we denote the two regimes as the para phase at low $D$ and the disordered phase at intermediate $D$. To check this interpretation, it is possible to look at the correlation in the imaginary time, defined as 
\begin{equation}
\label{eq:eq:imag_time_corr}
C_T(\tau\zeta)=\Big\langle \frac{1}{NP}\sum_i^N\sum_j^Pq^{(j)}_iq^{(j+\tau)}_i\Big\rangle.
\end{equation}
It is averaged over the simulation time, while $\tau$ is an integer running from $0$ to $P$, specifying the distance in imaginary time, bounded by $\beta=1 / k_bT$. In the para regime the particles do not distinguish between being right and left, and their position is randomly distributed in imaginary time. When the interaction is increased and the underlying ground state undergoes a displacive transition, a whole ring or sections of a ring become trapped in one of the minima and the correlation increases. Upon crossing the para-to-disorder crossover, individual particles are locked in the minima and correlation saturates, which is the hallmark of the local moment regime. Looking at the imaginary time correlation at half inverse temperature in the upper panel of Fig.~\ref{fig:Spin-freezing}, we see that indeed the correlation vanishes at low temperatures in the para regime and increases with temperature, due to the sharp decrease of de Broglie wavelength $\Lambda = \sqrt{2\pi \hbar^2/mk_b T}$, which determines the spread of the particle's ring. After the crossover line of local moment formation is crossed, the imaginary time correlation saturates and does not show temperature dependence any more, because thermal fluctuations can at most move the full particle or a large section of the ring. Indeed, the particles are already localized due to the interaction $D$, and displacing each individual bead becomes too expensive. We located the crossover by using a similar procedure to the one described by Werner \emph{et al.}\cite{Millis2008}. We compute the $C_T$ correlations at different temperatures and compare them to the one at $T=100$ K. By noticing that the ratio of correlation functions scales exponentially, we were able to precisely locate the critical value of $D$, where the ratio reaches unity. Indeed, the saturation point happens almost exactly at the value of $D$ where the R\'enyi entropy reaches its maximum (see Fig.~\ref{fig:Phase diagram}), unambiguously identifying the local moment formation point. The described behavior is clearly visible, if a system configuration is taken at random along the PIMD trajectory, and all the beads positions are mapped to $\{-1, 1\}$, i.e. left and right, with a sign function (see Fig.~\ref{fig:Spin-freezing-paths}). Ordering in imaginary time starts to appear around the QCP, when \emph{world-line} paths become much stiffer. 

Based on all these observations, we finally draw the converged phase diagram of the anharmonic chain in Fig.~\ref{fig:Phase diagram}. In a general 1D $\phi^4$ model we expect to see a gradual disappearance of the para regime, upon increase of the double well barrier. This should be replaced by the disordered regime so that eventually, in the limit of infinite double well barrier, the phase diagram resembles the one of the Ising model.

\begin{figure}[t]
\includegraphics[width=1\linewidth]{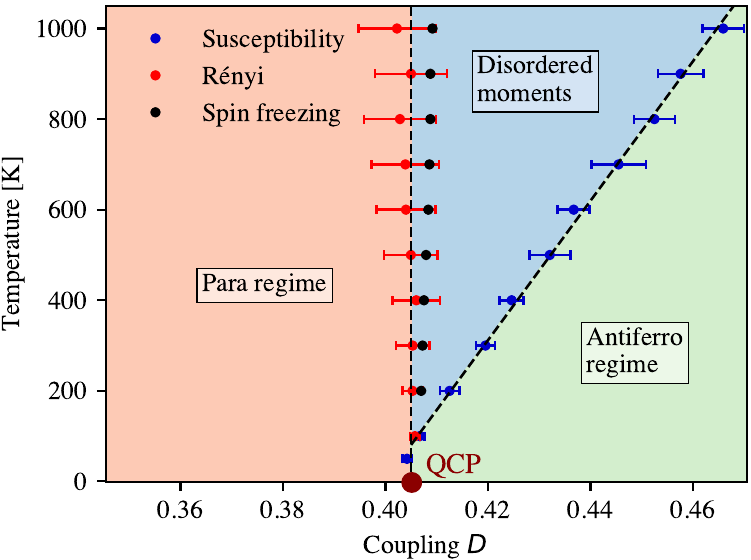}
\caption{Phase diagram of 1D an-harmonic chain in Eq.~\eqref{eq:ChainOfDoubleWellsPotential}. We observe three distinct regimes - para (\textit{red}), disordered (\textit{blue}) and antiferro (\textit{green}). The crossovers between these regimes are located by looking at the peak of the R\'enyi entropy (\textit{red} points), the ratio of the correlation functions in imaginary-time $C_T(\beta/2)$ of Eq.~\eqref{eq:eq:imag_time_corr} and Fig.~\ref{fig:Spin-freezing} (\textit{black} points, ``Spin freezing''), and the susceptibility $\chi$ of the order parameter in Eq.~\eqref{eq:Order paramether Gubernatis} and Fig.~\ref{fig:ClassicalScan} (\textit{blue} points, ``Global order''). Results are obtained at finite $\zeta$ and for $N=128$, and then extrapolated to $\zeta=0$, by shifting all points by $\delta D=0.01546$ to the right, as demonstrated for the R\'enyi entropy and susceptibility peaks (see Appendix~\ref{appendix:The scaling in zeta}). The error bars represent the width of the peaks at $95\%$ of their height.
\label{fig:Phase diagram}}
\end{figure} 

\section{Conclusion}
\label{sec:Conclusion}

We demonstrated that the R\'enyi entropy can be used to pin-point and classify phase diagrams. In the chain of anharmonic oscillators studied here, we successfully identified its quantum critical point and proved that the R\'enyi entropy, by successfully counting the number of available states, can also be used for the exploration of its finite-temperature phase diagram. 

The resulting phase diagram is exciting, because it shows features very similar to the ones of hydrogen rich materials. In particular, it is worth commenting on the apparent similarity between our 1D chain and the phase diagram of high-pressure water ice across the VIII-VII-X phase transitions, a region of its complex phase diagram that recently attracted significant attention\cite{Cherubini_2022}. Phase VIII of water ice is a crystalline phase, where a global network of hydrogen bonds is created. While at low-enough pressure (low density) the individual water molecules still have some mobility, as they can rotate and switch between 6 different configurations, phase VIII is a proton-ordered phase, with an order parameter analogous to our polarisation parameter. After the phase boundary to ice VII is crossed, protons can hop between neighbouring oxygen atoms. This is due to the increase of either temperature or pressure, and it is again analogous to the order-disorder crossover observed in our model. In realistic simulations and experiments, this is a first-order thermodynamic phase transition\cite{Pruzan2003,Reinhardt2022}. By further increasing the pressure and still following our analogy, the transition from ice VII to ice X can be interpreted as a quantum phase transition where quantum fluctuations destroy the local moments, created by the off-centered proton positions with respect to the two flanking oxygen atoms. Therefore, phases VII and X are represented by the disordered quantum critical phase and the para phase of our model, respectively. Based on our analogy, a direct transition from phase VIII to phase X is then expected at low temperature. This analysis is however a very crude simplification, and a full quantum mechanical treatment of the realistic system should be carried out. With the method developed and presented in this work, this is certainly within our reach. Nevertheless, VIII-VII-X water ice is only one of the practical realizations of the more general $\phi^4$ model. Our R\'enyi entropy approach can thus be fruitfully applied to a large variety of quantum critical systems.

We have shown that the R\'enyi entropy can be used to detect local moments formation in many-body quantum systems. This owes to the fact that the R\'enyi entropy counts the average number of occupied states in the system. As a result, at sufficiently low temperature, the local moments formation leads to an abrupt change in entropy, which can be seamlessly connected with the entropy divergence at the zero-temperature QCP. Therefore, the R\'enyi entropy can be used to extensively characterize temperature-dependent phase diagrams of quantum systems, and its sensitiveness to emergent local symmetry breakings makes it a precious tool to localize elusive phase transitions that otherwise would be very hard to detect. 

\begin{acknowledgments}

We thank GENCI for the computational resources provided through the allocation A0130906493, where the main calculations have been performed. We thank the support of the HPCaVe computational platform of Sorbonne University. We are grateful for the environment provided by ISCD and its MAESTRO junior team. This  work  was partially supported  by  the  European  Centre of Excellence in Exascale Computing TREX-Targeting Real Chemical Accuracy at the Exascale, funded by the European Union’s Horizon 2020 Research and Innovation program under Grant Agreement No.~952165.

\end{acknowledgments}
\appendix

\section{Numerical details}
\label{appendix:Numerical details}

The value of the damping constant $\gamma_0$ was fixed to $\gamma_0=0.005855$ atomic units (a.u.). On the other hand, for the Born-Oppenheimer (BO) Liouvillian, $\gamma^{BO}$ has been set to zero, because the BO forces of our model are deterministic and $\gamma_0 \neq 0$ was enough to thermalize the system. A typical simulation was split in blocks of $10000$ time propagation steps, and usually $140$ to $400$ blocks were evaluated. Time step was calculated with the formula $\Delta t = 0.25 D/0.334$ fs, with $D$ defined in Eq.~\eqref{eq:ChainOfDoubleWellsPotential}, and varied form $0.25-0.375$ fs. Final simulation times were therefore ranging from $0.1$ ns to $0.28$ ns. This long simulation times were needed in the vicinity of the QCT, due to symmetry breaking, after which two distinct copies of the system got trapped in distinct global minima. In this case the problem was relieved by introducing a SWAP move (see Appendix~\ref{appendixA:Accelerating MD sampling with SWAP operator}). However, longer simulation times were still needed, to reach low error estimates.

Temperature was varied from $50$ K to $1000$ K. This resulted in the range of $\beta$ from $315$ Ha$^{-1}$ to $6313$ Ha$^{-1}$, where Ha is atomic unit of energy. The maximum number of beads used was $50$. For the numerical integration over the regularized thermodynamic integration path we used $10$ integration steps.

\section{Accelerating MD sampling with SWAP move}
\label{appendixA:Accelerating MD sampling with SWAP operator}

As is often the case, the simulations based on PIMD can get stuck in minima separated by the barriers larger than the size of thermal fluctuations. This is particularly true when calculating the R\'enyi entropy in the models with broken symmetry, like for example in the double well potential. In such case in the regime with large barriers different replicas belonging to the ensemble get stuck in either the same, or different minima. In order to get the correct estimation of entropy both configurations must be sampled, which becomes exponentially slower with the size of the barrier.

To facilitate the sampling, a SWAP move can be introduced, especially if minima of the system are related by symmetry. The SWAP move flips all the particles in one of the replicas from one minimum to another. During the simulation this move is proposed and accepted with the probability given by the Boltzmann distribution. The main change in energy comes from the boundary conditions connecting the two ensembles. Note that the velocities can be rotated in any fashion, since the norm of them stays unchanged.

Given the symmetry of the anharmonic chain, the SWAP was done by multiplying all particle positions by $-1$. Using this additional SWAP move was enough to preserve ergodicity in our simulations. We believe that such scheme can be used also in systems with larger symmetries, such as higher dimensional chain of anharmonic oscillators.

\section{Method testing}
\label{appendixB:Tests of the method}

As a test of performance of our algorithm, we used a system of two coupled harmonic oscillators. The model is described by the Hamiltonian in Eq.~\eqref{eq:NparticleHamiltonian} where the potential is given by
\begin{equation}
    \label{Appeq:CHO_potential}
    V(q_1, q_2) = \frac{m_1\omega_1^2}{2}q_1^2+\frac{m_2\omega_2^2}{2}q_2^2 - m\omega_1\omega_2\Gamma q_1q_2
\end{equation}
The Hamiltonian is bounded from bellow only for $|\Gamma|<1$ where there exist a global minimum at the point $(x_1,x_2)=(0, 0)$. Since the Hamiltonian comes as a quadratic form, the coordinate system can be rotated in such a way that it actually describes two non interacting Harmonic oscillators
\begin{eqnarray}
y_1 = \cos\delta \,q_1 - \sin\delta\, q_2\nonumber\\
y_2 = \sin\delta\, q_1 + \cos\delta\, q_2,
\label{Appeq:CoupledHarmonicHam_rotation}
\end{eqnarray}
with the angles expressed as\cite{Makarov2018}
\begin{eqnarray}
\epsilon = \frac{\omega_2^2 - \omega_1^2}{2\omega_1\omega_2\Gamma},\\
\tan\delta = \frac{\epsilon}{\|\epsilon\|} \sqrt{\epsilon^2 + 1} - \epsilon,
\label{Appeq:CoupledHarmonicHam_angles}
\end{eqnarray}
and the new frequencies given by $\tilde{\omega}_i^2 = \omega_i^2 \pm \omega_1\omega_2\Gamma\tan\delta$. Each of these oscillators has a well known density matrix $\rho_0$, given by a Gaussian distribution\cite{Feynman1972}
\begin{eqnarray}
\rho_0(y_i, y_i';\beta)=\sqrt{\frac{2\xi_i-\psi_i}{\pi}}e^{-\xi_i(y_i^2+y_i'^2)+\psi_iy_iy_i'},
\label{Appeq:DensityOfOneOscillator}
\end{eqnarray}
with the new parameters $\xi_i,\psi_i$ defined as
\begin{eqnarray}
\xi_i =& \frac{m\tilde{\omega}_i}{2\hbar}\coth(\hbar\tilde{\omega}_i\beta),\\
\psi_i =& \frac{m\tilde{\omega}_i}{\hbar}\frac{1}{\sinh(\hbar\tilde{\omega}_i\beta)}.
\label{Appeq:xi_and_psi}
\end{eqnarray}
The density matrix of the full system is therefore given by $\rho=\rho_0(\tilde{\omega_1},y_1)\rho_0(\tilde{\omega_2},y_2)$. In order to perform traces over original degrees of freedom $q_{1,2}$, we perform back rotation, by using Eq.~\ref{Appeq:CoupledHarmonicHam_rotation}. Now all the traces of all the powers of the density matrix and of the reduced density matrix can be evaluated analytically with multivariate Gaussian integrals. 

For the full R\'enyi entropy of order $\alpha$, we need to evaluate the trace of the density matrix to the integer power $\alpha$, $\rho^{\alpha}$. It can be again expressed as a product of Gaussian distributions, given by the tridiagonal quadratic forms $\rho_0^{\alpha}(y_i;\beta)=\exp(-\textbf{y}_i^TA_i\textbf{y}_i/2)$ of dimension $\alpha\times\alpha$. The new vector $\textbf{y}_i$ contain coordinates $y_i$ belonging to separate replicas of a corresponding free mode, and the matrix coupling them is given by a tridiagonal matrix, with terms $4\xi_i$ on the main diagonal and terms $-\psi_i$ as sub- and supra- diagonals. There are also two additional $-\psi_i$ terms added to the off-diagonal corners. To procede, we note that the multivariate Gaussian integral of a quadratic form equals the inverse of the square root of its determinant. Therefore, to evaluate the R\'enyi entropy of the full complex it is enough to evaluate the two determinants $|A_{i}|$, so that
\begin{equation}
    \label{Appeq:FinalFullCHO}
    S^{\alpha} = \frac{1}{1 - \alpha}\log\Pi_i\left[\Big(\frac{2\xi_i-\psi_i}{\pi}\Big)^{\frac{\alpha}{2}}\frac{(2\pi)^{\frac{\alpha}{2}}}{|A_i|^{\frac{1}{2}}}\right].
\end{equation}
For example, when $\alpha=2$ it equals to
\begin{eqnarray}
\label{eq:CoupledHarmonicHam_FullEntropy}
S^2=-\log(\tanh(\hbar\tilde{\omega}_1\beta/2)\tanh(\hbar\tilde{\omega}_2\beta/2)),
\end{eqnarray}
demonstrating that indeed the system has a non-degenerate ground state, since $S^2=0$ in the limit of $T=0$.

Completely analogous steps have to be done also for the reduced density matrix. By integrating over $q_1$ we obtain the reduced density matrix of the second oscillator, defined as 
\begin{eqnarray}
\rho_2(q_2, q_2';\beta) = B\exp\bigg\{q_2^2(\Phi^2-\Xi)\nonumber\\+q_2q_2'(2\Phi^2+\Psi)+q_2'^2(\Phi^2 - \Xi)\bigg\},
\label{Appeq:CoupledHarmonicHam_RedDensitySol}
\end{eqnarray}
where the new parameters are given through the prescriptions below.

\begin{eqnarray}
\Xi =& \xi_1\sin^2\delta + \xi_2\cos^2\delta,\\
\Psi =& \psi_1\sin^2\delta + \psi_2\cos^2\delta,\\
\Phi =& \frac{2\xi_1 - \psi_1 - 2\xi_2 + \psi_2}{2\sqrt{\frac{2\xi_1 - \psi_1}{\sin^2\delta}+\frac{2\xi_2 - \psi_2}{\cos^2\delta}}},\\
B =& \frac{m}{\pi\hbar}\sqrt{\frac{\pi\tilde{\omega}_1\tilde{\omega}_2\tanh(\hbar\tilde{\omega}_1\beta/2)\tanh(\hbar\tilde{\omega}_2\beta/2)}{(2\xi_1 - \psi_1)\cos^2\delta+(2\xi_2 - \psi_2)\sin^2\delta}}
\label{eq:CoupledHarmonicHam_RedDensityConstants}
\end{eqnarray}
Finally, to get the trace of any power of the reduced density matrix, one can notice that the density matrix of integer power $\alpha$, $\rho_2^{\alpha}(q_2,q_2:\beta)$, is again a Gaussian distribution, given by the tridiagonal quadratic form $\exp(-\textbf{q}_2^TA\textbf{q}_2/2)$ of dimension $\alpha\times\alpha$. Now the vector $\textbf{q}_2$ contains coordinates $q_2$ belonging to separate replicas. The matrix $A$ has $4(\Phi^2 - \Xi)$ on the main diagonal and terms $(2\Phi^2+\Psi)$ as sub- and supra- diagonals. There are also two additional $(2\Phi^2+\Psi)$ terms added to the off-diagonal corners. By using the same formula for the multivariate Gaussian integral of a quadratic form, we get the result for the R\'enyi entropy of subsystem 2:
\begin{equation}
    \label{Appeq:FinalPartialCHO}
    S^{\alpha}_2 = \frac{1}{1 - \alpha}\log\Big[B^{\alpha}\frac{(2\pi)^{\frac{\alpha}{2}}}{|A|^{\frac{1}{2}}}\big)\Big].
\end{equation}
which in the case of $\alpha=2$ reduces to a much simpler expression
\begin{eqnarray}
S^2_2=-\log(B^2\frac{\pi}{\sqrt{4(\Xi-\Phi^2)^2-(2\Phi^2+\Psi)^2}}).
\label{eq:CoupledHarmonicHam_EntanglementEntropy}
\end{eqnarray}
A similar expression can be found also for the first oscillator, by everywhere replacing $\cos\delta$ with $\sin\delta$ and \textit{vice versa}.

The system represents an ideal continuous model with entanglement and can be used as a bench-marking model for algorithms evaluating entanglement entropy. The level of entanglement in this system can be very large and is diverging upon approaching the limit of $\Gamma = 1$\cite{Zhou2020}. In the following tests we used the value that is very close to the critical point $\Gamma=0.94868$, but distant enough to preserve good convergence of the MD algorithm. 

\begin{figure}[!ht]
\includegraphics[width=1\linewidth]{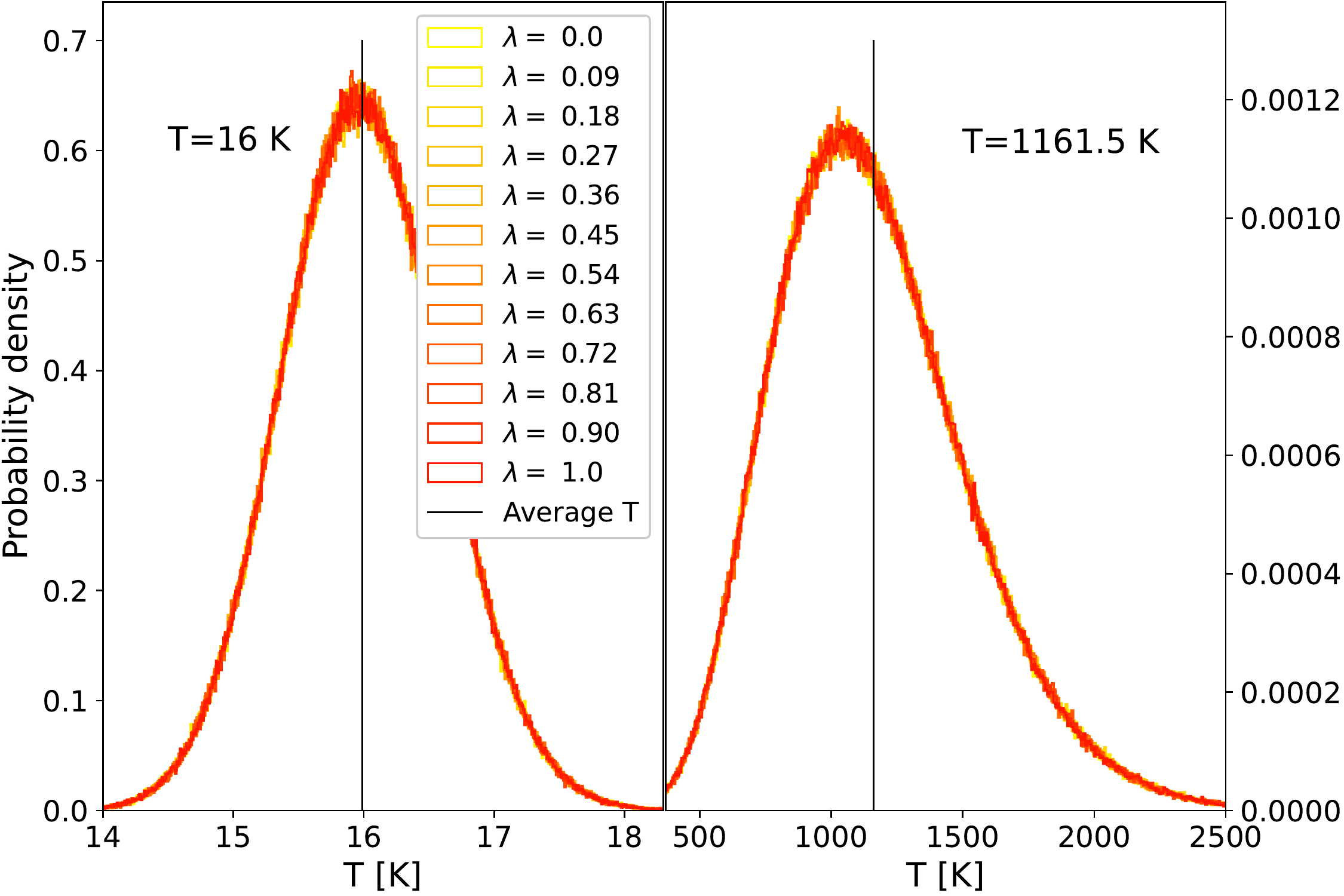}
\caption{Temperature distribution along the integration path evaluating the entropy of Harmonic oscillator with $\omega=0.01$, when coupled to another with $\omega=0.004$, through $\Gamma=0.94868$, defined in Eq.~\ref{Appeq:CHO_potential}.
\label{fig:TemperatureDistribution}}
\end{figure}  

First we used the model to confirm that the coupling to the thermal bath is done correctly with the prescription in Eq.~\eqref{eq:GammaMatrixElementModified}. We tested the coupling to the thermal bath by looking at the distribution kinetic energy, expressed as an instantaneous temperature, along the trajectories for different ensembles of coupled Harmonic oscillator along the path in Eq.~\eqref{eq:Path}, corresponding to the evaluation of the entropy of one subsystem. In this case the friction changes from particle to particle and for different values of the coupling constant.  As can be seen in Fig.~\ref{fig:TemperatureDistribution}, all trajectories are correctly thermalized, validating the algorithm presented in Sec.~\ref{subsec:Path Integral Ornstein-Uhlenbeck Dynamics (PIOUD)}.

\begin{figure}[h!]
\includegraphics[width=1\linewidth]{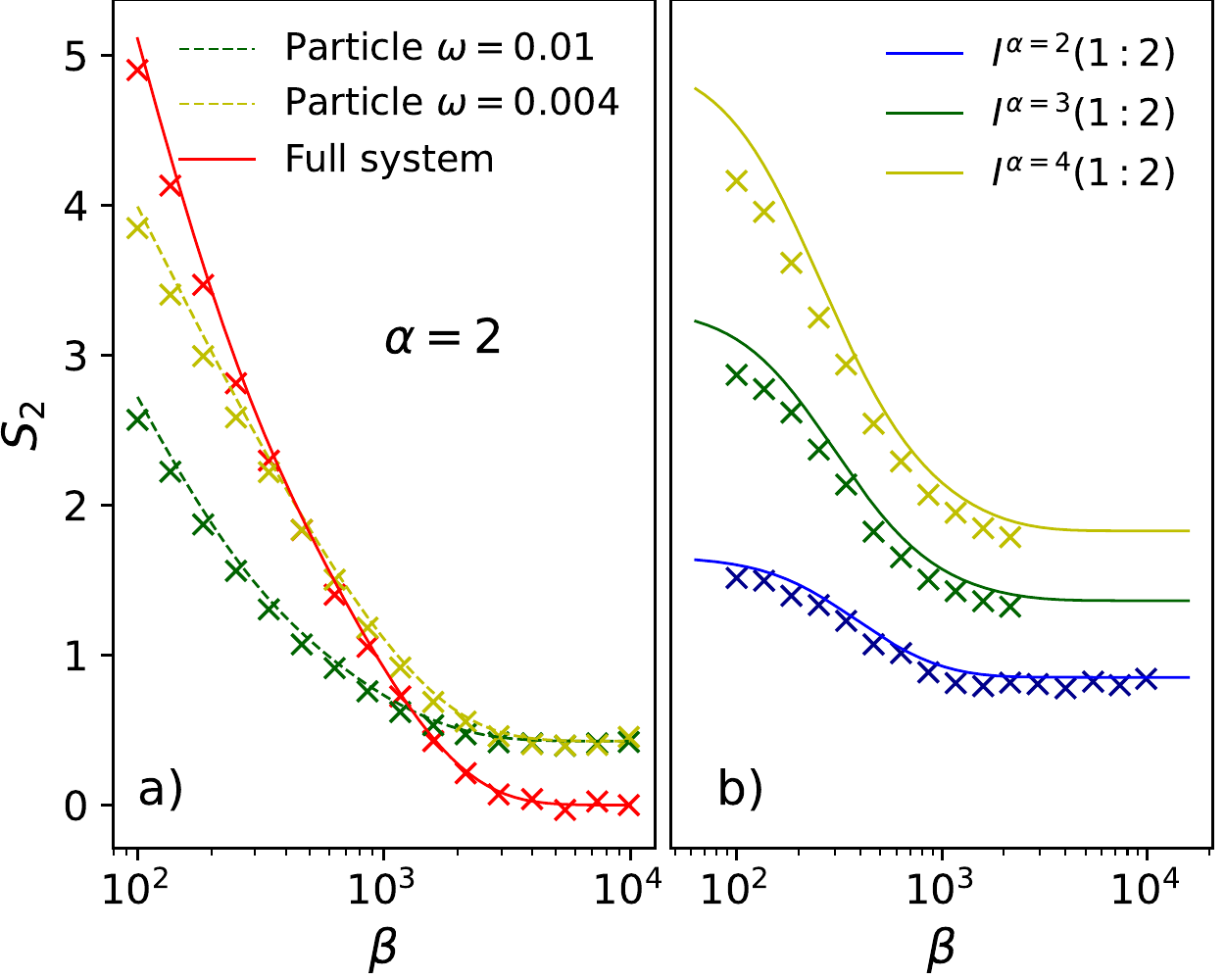}
\caption{Evaluation of R\'enyi entropy of harmonic oscillator with $\omega=0.01$ and an oscillator with $\omega=0.004$ that are coupled through $\Gamma=0.94868$, defined in Eq.~\ref{Appeq:CHO_potential}. a) Entropy at different temperatures. b) R\'enyi Mutual information (see Eq.~\eqref{eq:RenyiMutualInformation}) of higher orders $\alpha=2,3,4$. Numerical results (crosses) are compared to the exact analytical results (line).\label{fig:CHOEntropy}}
\end{figure} 

\begin{figure*}[t]
\centering
\includegraphics[width=\linewidth]{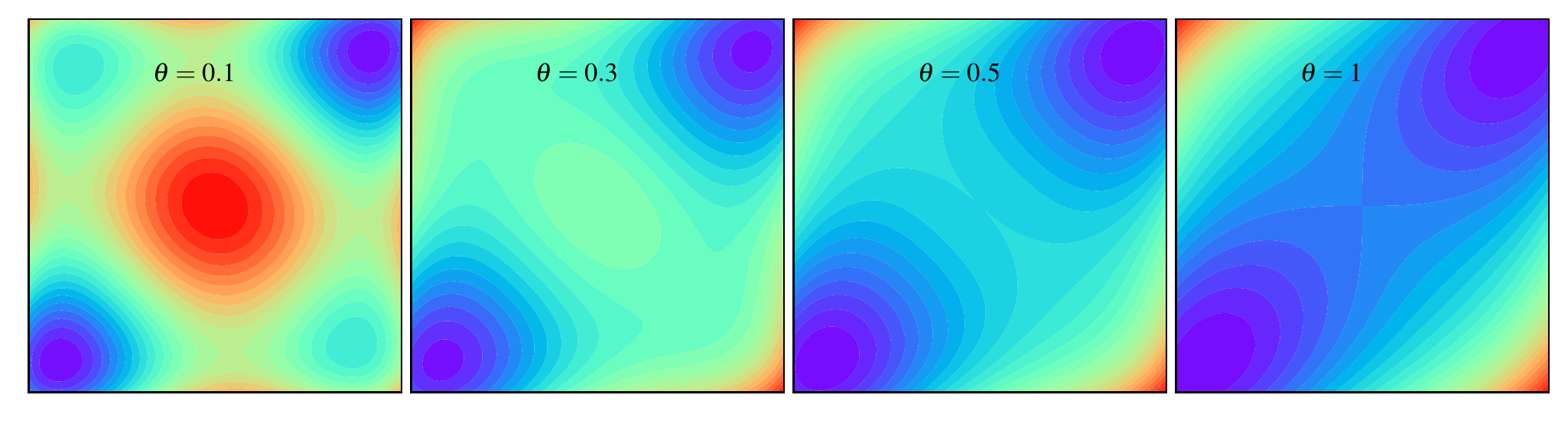}
\caption{Classical potential as a function of parameter $\theta$. For simplicity, only a classical potential of two particles is shown, with all the other parameters fixed to one $m=\omega=\lambda=1$.\label{Appfig:Potential}}
\end{figure*}

We then compared numerical and analytical results for the entropy, and tested the performance of the algorithm in evaluating higher orders of the R\'enyi entropy. At higher orders of R\'enyi entropy the test was performed by comparing the simulation with the exact result for the R\'enyi mutual information. It can be expressed as
\begin{equation}
\label{eq:RenyiMutualInformation}
I^{\alpha}(1:2)=S^{\alpha}_1+S^{\alpha}_2-S^{\alpha},
\end{equation}
where the lower index specifies which oscillator was chosen as a subsystem in the evaluation of R\'enyi entropy of order $\alpha$. This quantity is useful because it cancels out the correlations of the full system and really quantifies the correlations between the particles. However the quantity is really related to the correlation only in the limit of $\alpha\to1$, when it cannot be negative. In the system of coupled harmonic oscillators the mutual information of integer order is never negative and follows a similar curve as $\alpha=1$. This suggests that it is related to the actual correlation. 

It can be seen from the Fig.s~\ref{fig:CHOEntropy} that the method agrees well with the theoretical result, with a small bias due to numerical integration. The method can cover a very large span of temperatures, thanks to the choice of optimal integration path. However, the bias increases slightly with temperature, see Fig.~\ref{fig:CHOEntropy}a, and with increasing  order $\alpha$, see Fig.~\ref{fig:CHOEntropy}b. The bias can be combated by re scaling the integration parameter $\lambda$, but for our purposes this is not needed, since in this work we analyse low temperature behaviour of the second order R\'enyi entropy. 

\section{The \texorpdfstring{$\phi^4$}{φ4} model}
\label{appendix:phi4 model}

The model with the potential in Eq.\eqref{eq:DoubleWellPotential}, is described by the Hamiltonian
\begin{equation}
\label{Appeq:DoubleWellHamil}
\hat{H}(\hat{\textbf{p}}, \hat{\textbf{q}}) = \sum_i^N \frac{\hat{p}_i^2}{2m}+\sum_i^N \frac{\theta}{2}(\hat{q}_i-\hat{q}_{i+1})^2-m\omega^2\hat{q}_i^2+\lambda\hat{q}_i^4.
\end{equation}
Considering two particles, when $\theta < m\omega^2 / 2$ the classical potential features four minima. Two global minima are located at
\begin{equation}
\label{Appeq:GlobalMinima}
q^m_{0}=q^m_{1}=\pm\sqrt{\frac{m\omega^2}{2\lambda}},
\end{equation}
and two other minima correspond to
\begin{equation}
\label{Appeq:OtherMinima}
    q^m_{0}=-q^m_{1}=\pm\sqrt{\frac{m\omega^2}{2\lambda}-\frac{\theta}{\lambda}},
\end{equation}
which become imaginary, when $\theta > m\omega^2 / 2$. On Fig.~\ref{Appfig:Potential} we show some possible potential energy surfaces,for the case of two particles. In this work, we have set $\theta=m\omega^2$ which leaves only two minima as discussed in the main text. In doing so we defined new parameters $m'=m\lambda$ and $D=\sqrt{m'}\omega$, leading to the Hamiltonian of Eq.~\eqref{eq:DoubleWellPotential} with an imaginary coupling, that is corresponding to the ferroelectric situation. 

There exist many different parametrisations of the model. For making comparison with previous results, we list the mappings from our parametrisation to some others. In the works\cite{Savkin2002, Rubtsov2001} the Hamiltoniain is written in terms of parameters $a, \sigma$ and $d$, such that
\begin{equation}
\label{Appeq:DoubleWellHamilRubtsov}
\hat{H}(\hat{\textbf{p}}, \hat{\textbf{q}}) = \sum_i^N \frac{\hat{p}_i^2}{2m_i}+\sum_i^N \Big(2d - \frac{a}{2}\Big)\hat{q}_i^2 + \frac{a}{4}\hat{q}_i^4-\sigma\hat{q}_i\hat{q}_{i+1}.
\end{equation}
With a simple comparison we can see that then $\lambda=a/4$, $\theta=\sigma$ and $\omega = \sqrt{\frac{3a}{4m}-2\frac{d}{m}}$. This Hamiltonian is particularly useful for exploring the limit of the Ising model ($a\to\infty$), where particles are strongly localized to positions around $-1$ and $1$. The mass causes them to tunnel and is analogous to the strength of the transverse magnetic field in the Ising model. 

When comparing our predictions for the critical coupling $D^*$ with the work of Wang \emph{et al.}\cite{Wang1994}, the following correspondence is used. They show that the model can be recast to depend on only two parameters $\kappa$ and $\epsilon$. The two parameters represented in the Hamiltonian formulation appear as
\begin{equation}
\label{Appeq:ActionGubernatis}
S[\phi]=\epsilon\sum_i^N\Big[\frac{\hat{p}^2}{2\kappa}+\frac{\kappa}{2}\big(\hat{q}_i-\hat{q}_{i+1}\big)^2+\frac{1}{4}(\hat{q}_i^2-1)^2\Big],
\end{equation}
and can be expressed in terms of our parameters as $\omega = (\epsilon\lambda / m^2)^{1/3}$ and $\kappa=\theta / 2m\omega^2$. From our analysis we can see that the last one controls the type of the transition and the number of minima, while the first one controls the strength of quantum effects. The larger the value of $\epsilon$, closer we are to the classical limit. By making the restriction $\theta=m\omega^2$, we choose the value $\kappa'=1/2$ and vary only $\epsilon$, in this case $\epsilon'=m^2\omega^3/\lambda=\sqrt{m'}D^3$, which we use to compare our results.

\section{Scaling in \texorpdfstring{$\zeta$}{ζ}}
\label{appendix:The scaling in zeta}

\begin{figure}[h!]
\includegraphics[width=\linewidth]{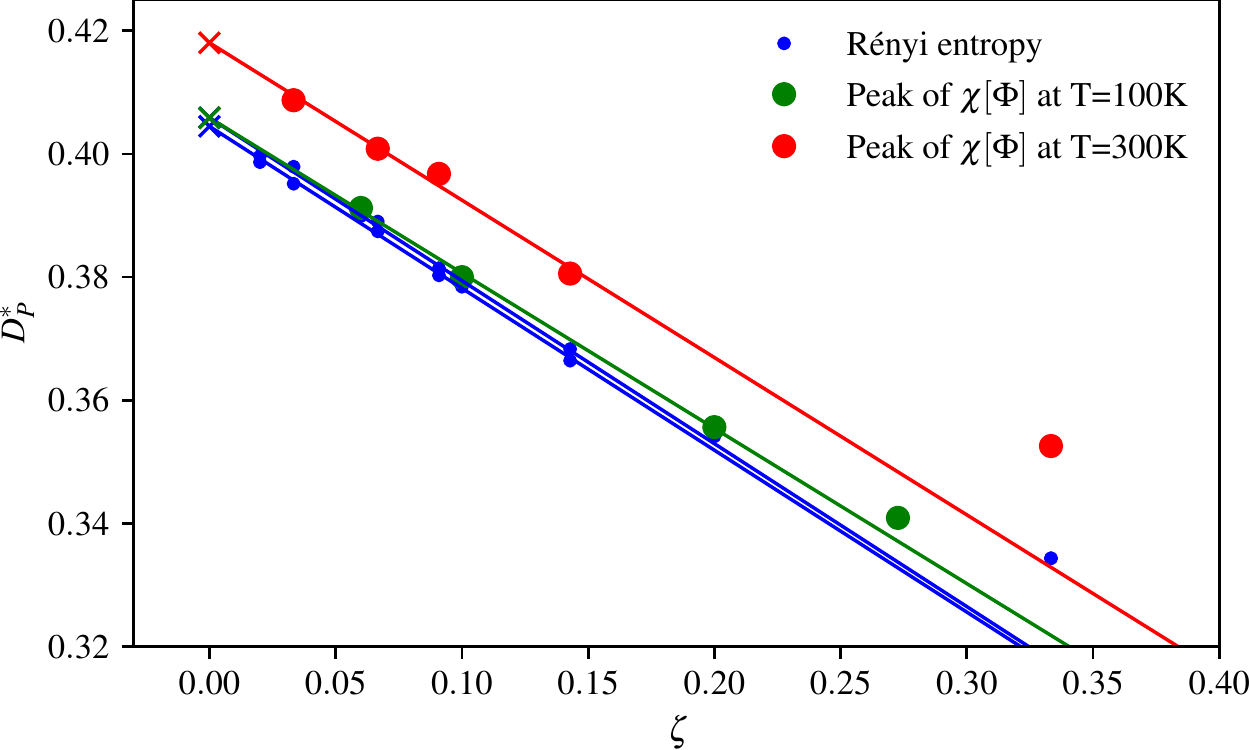}
\caption{$\zeta = 300 / PT$ dependence of the peaks of the susceptibility of the order parameter, introduced in Eq.\eqref{eq:Order paramether Gubernatis}. \textit{Red} points correspond to the susceptibility at $300$K and \textit{green} points to the susceptibility at $100$K. For comparison we show also the positions of the peaks of the R\'enyi entropy in \textit{blue}, for both temperatures and two subsystem sizes. The linear interpolation shows that the peaks of all the quantities drift as a function of $\zeta$ with the same rate.
\label{Appfig:Scaling_susc}}
\end{figure}

We observed that the position of the peak of the R\'enyi entropy strongly depends on the size of the imaginary-time step $\zeta$ (Fig.\ref{fig:EntropyScan}). This dependence appears linear in $\zeta$, which is large, compared to the dependence of the partition function - $\zeta^3$ - given by the trotterization. To determine the full effect of $\zeta$ on the phase diagram, Fig.~\ref{fig:Phase diagram} we further studied the scaling of the susceptibility upon the discretisation in the imaginary-time. In Fig.~\ref{Appfig:Scaling_susc} we show the position of the susceptibility peak for different temperatures and imaginary-time steps. The maximum of the R\'enyi entropy is shown on the same plot. It can be seen that, at these temperatures, the relative positions of these maxima converges very early, while the absolute position converges, linearly, as the position of the peak of the R\'enyi entropy. Since we observed no temperature dependence on the convergence rate, we expect this approximation to hold also at slightly higher temperatures, covered in Fig.~\ref{fig:Phase diagram}. Therefore the phase diagram in the limit of zero imaginary-time can be obtained, by just recording the positions of the peaks at sufficiently small imaginary-time step and then shifting all of them with the linear fit in Fig.~\ref{Appfig:Scaling_susc}.

\bibliography{PRB_citations}

\end{document}